\documentclass[aps,prc,eqsecnum,showpacs,twocolumn,letterpaper,amssymb,amsmath,nofootinbib,floatfix]{revtex4-1} %elsNot
\usepackage{bm} %elsNot
\usepackage{graphicx} %elsNot
\usepackage{array} %elsNot
\usepackage{tabularx} %elsNot
\usepackage{dcolumn} %elsNot
\usepackage{ifthen} %elsNot

\begin{document}
\def\sgn{\mbox{sgn}}

\newcounter{minutesToday}
\newcounter{hours}
\newcounter{hmins}
\newcounter{minutes}
\setcounter{minutesToday}{\time}
\setcounter{minutes}{\time}
\setcounter{hours}{\time}
\divide \value{hours} by 60
\setcounter{hmins}{\value{hours}}
\multiply \value{hmins} by 60
\addtolength{\value{minutes}}{-\value{hmins}}
\def\HHMM{\ifthenelse{\theminutes<10}{\thehours:0\theminutes}{\thehours:\theminutes}}

%els\begin{frontmatter} %els

  \title{Muon Reconstruction in the Daya Bay Water Pools}
  \author{R.W.~Hackenburg} %elsNot
  \affiliation{Physics Department, Brookhaven National Laboratory, Upton, NY 11973}  %elsNot

%els  \def\BNL{1} %els
%els  \author[\BNL]{R.W.~Hackenburg} %els
%els  \address[\BNL]{Brookhaven~National~Laboratory, Upton, New York, USA} %els

\date{\today\qquad\HHMM\qquad DRAFT}

\begin{abstract}
Muon reconstruction in the Daya Bay water pools would serve to verify the simulated muon fluxes and offer
the possibility of studying cosmic muons in general.
This reconstruction is, however, complicated by many optical obstacles
and the small coverage of photomultiplier tubes (PMTs) as compared to other large water Cherenkov detectors.
The PMTs' timing information is useful only in the case of direct, unreflected Cherenkov light.
This requires PMTs to be added and removed as an hypothesized muon trajectory is iteratively improved,
to account for the changing effects of obstacles and direction of light.
Therefore, muon reconstruction in the Daya Bay water pools does not lend itself to a general fitting procedure
employing smoothly varying functions with continuous derivatives.
Here, an algorithm is described which overcomes these complications.
It employs the method of Least Mean Squares to determine an hypothesized trajectory from the PMTs' charge-weighted positions.
This initially hypothesized trajectory is then iteratively refined using the PMTs' timing information.
Reconstructions with simulated data reproduce the simulated trajectory to within about $5^\circ$ in direction
and about 45~cm in position at the pool surface,
with a bias that tends to pull tracks away from the vertical by about $3^\circ$.
\end{abstract}

%els  \begin{keyword} %els
%els    % keywords here, in the form: keyword \sep keyword
%els    Neutrinos \sep Water Shield \sep Cosmic Rays \sep Muons \sep Underground \sep Reconstruction
%els    % PACS codes here, in the form: \PACS code \sep code
%els    \PACS 07.77.Ka \sep 13.88.+e \sep 29.27.Hj \sep 41.75.Fr
%els  \end{keyword} %els

%els \end{frontmatter} %els
\maketitle %elsNot

\section{Detector Synopsis}
\label{sec1}
 \begin{figure*}
 \includegraphics[width=\textwidth]{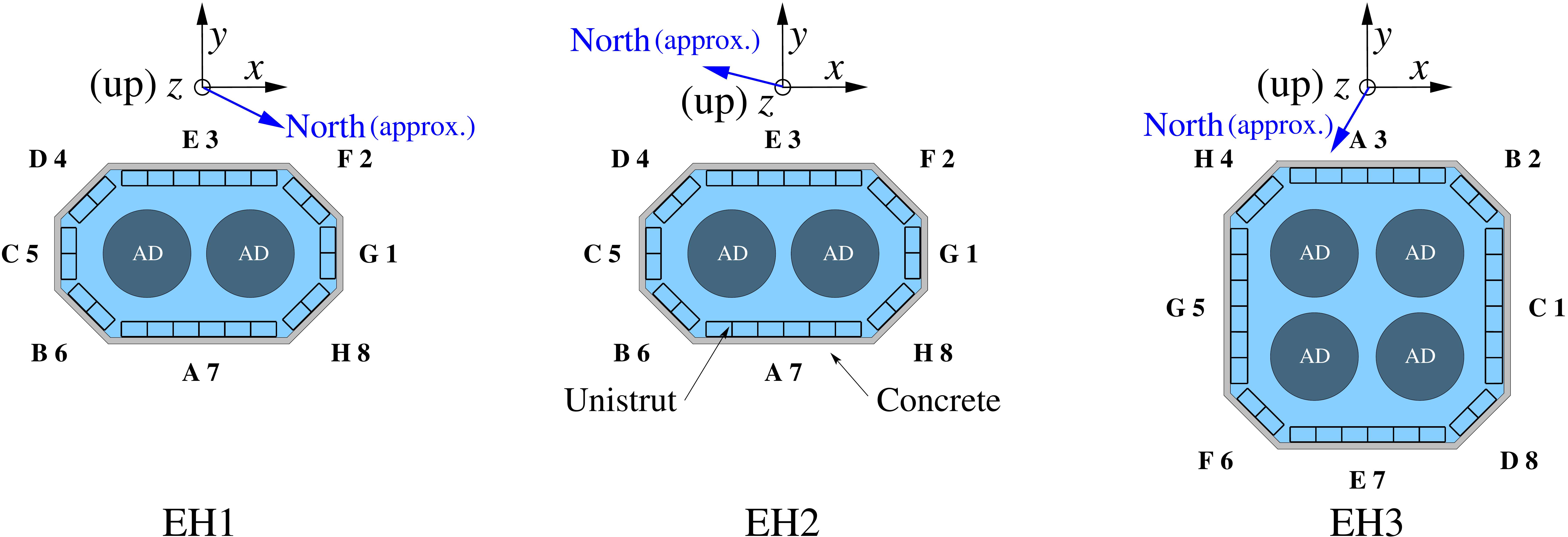}
 \caption{\label{Fig1} Sketches (not to accurate scale) showing the water pools in the three experimental halls.
  The pools are all 16~m long in $x$ and 10~m deep.
  EH3 is 16~m wide in $y$, while EH1 and EH2 are 10~m wide.
  The corner walls are at $45^\circ$ to their neighboring walls.
  Each pool has eight walls, designated with letters A-H in engineering documentation,
  and with numbers 1-8 in software (which designates the floor as ``wall 9'').
  Note that the correspondence between these two sets of labels is different in EH3 than in EH1 and EH2.
  The pools' inner and outer optical zones are separated from each other by highly reflective Tyvek covering the Unistrut frame.
  The inner zones of EH1 and EH2 are each instrumented with 121 PMTs, while their outer zones each have 167,
  for a total of 288 in each hall.
  The inner zone of EH3 has 160 PMTs and its outer zone has 224, for a total of 384.
  The ADs are all 5~m in diameter, 5~m high, and 2.5~m above the bottom of the pool.
  An RPC array, not shown, covers each pool.
  In each hall, the origin is at the center of the designed pool surface ($z$ is negative throughout the pool).
  % The true water surface is about 10~cm below that.
}
 \end{figure*}

The Daya Bay experiment~\cite{1stOsc,2ndOsc}
employs eight identically-designed antineutrino detectors (AD)
to observe the antineutrino flux from six nearby nuclear reactors.
The ADs reside in three separate experimental halls (EH)
with two ADs in each of the two near halls and four in the one far hall (Fig. \ref{Fig1}).
The ADs are immersed in ultrapure water pools, which are instrumented with photomultiplier tubes (PMTs).
The pools serve both as shields for the ADs against natural, low-energy radiation,
and also as water Cherenkov detectors~\cite{MuonPaper}.
The pools are each divided into inner and outer water shields (IWS and OWS),
separated by highly reflective Tyvek and very nearly optically isolated from each other.
The OWS comprises the outer 1~m of each pool's sides and bottom,
but does not cover the top.
The IWS and OWS are each populated with 20~cm photomultiplier tubes (PMT),
which detect the Cherenkov light produced by the passage
of a relativistic charged particle through the water.
Being underground, such a particle can safely be assumed to be a muon, where EH1, EH2, and EH3 have overburdens of
250, 265, and 860~mwe (meters water equivalent)~\cite{MuonPaper}.
The PMTs are a combination of new Hamamatsu PMTs and older EMI PMTs recycled from the MACRO experiment~\cite{macro}.
A rectangular array of resistive plate chambers (RPC) covers each pool,
with a pair of small RPC panels about 2~m above the main RPC array forming the RPC Telescope.

Each AD, IWS, OWS, and RPC is an independent detector subsystem.
There are a total of five subsystems each in EH1 and EH2 (AD1, AD2, IWS, OWS, RPC),
and seven in EH3 (AD1-AD4, IWS, OWS, RPC).
ADCs and TDCs record the charge and time of each PMT that exceeds a threshold of about 0.25~pe (photoelectrons).
The triggering and readout electronics of the PMT subsystems (ADs, IWS, and OWS)
are all identical~\cite{ADpaper,LocalTriggerBoard,DAQarch}.
Each PMT subsystem is independently triggered by its own energy sum and multiplicity triggers,
determined only by the subsystem's own PMTs and without regard to the other subsystems.
Because of the processing time of the trigger logic,
the trigger occurs $\sim 1$~$\mu{\rm sec}$ later than the recorded PMTs' times.
The TDC values are recorded in such a fashion that they are negative, with the latest times the most negative.
It is convenient to reverse this and shift the times positive, so that the earliest recorded times are
a little above zero, with the latest times appearing as the most positive.
To this end, the raw TDC values are subtracted from a fixed and somewhat arbitrary constant,
which is the same for all events and PMTs, and for all detector subsystems.
Thus, the reversed and shifted PMT time is
\begin{equation}
\label{1.1}
\tau \equiv (1100 - \mbox{TDC})\times 1.5625\,\mbox{ns}\, .
\end{equation}
Calculations involving the times are all differences, wherein the constant 1100 is always canceled.
These ADCs and TDCs record multiple threshold crossings in a given readout from a trigger,
but only the first such crossing, or hit, from each PMT is used here.
The RPC subsystem is triggered when at least three of the four layers in an RPC module has a pulse.
The separate subsystem readouts are combined into {\em events}.
Not all events contain readouts from all detectors, and some consist of readouts from single detectors.

Each PMT is characterized by a gain $\mu$ and a timing offset $t_0$, stored in a database and indexed by date and time.
The gains are calibrated on a continuous basis using dark noise previous to each readout.
The timing offsets correct for fixed differences in the timing of each PMT,
caused by individual variations in PMT transit delay,
signal propagation delay in the cable and electronics, etc.,
so that any two or more PMTs
seeing a short pulse of direct light from an equidistant source would all have the same corrected time $\tau - t_0$,
to within an error of about 2~ns.
There is also a small correction made for the variation in timing due to pulse height,
referred to as a {\em time walk} or {\em time slewing} correction,
which is subtracted from the time to make small signals' times earlier.
For 2~pe, the smallest signals used in track reconstruction, this correction is 6.8~ns,
but it falls rapidly to 0.6~ns at 20~pe and becomes negligible above 40~pe.

A set of calibration LEDs were installed in the pools to determine the PMT time offsets, among other uses,
but these LEDs began to fail early in the course of the experiment,
leaving ever-widening gaps in the set of PMTs which are illuminated by direct light from an LED.
As will be demonstrated, these offsets can be obtained from an analysis of a large sample of muon data.

\section{Initial Trajectory}
\label{sec2}
Absent any significant magnetic field, and considering only high energy cosmic muons,
a muon trajectory is a straight line.
In the local coordinate system of a hall (Fig. \ref{Fig1}),
a muon trajectory is specified separately in the $xz$ and $yz$ planes in terms of the common $z$ coordinate
and the four parameters $x_0$, $x' \equiv dx/dz$, $y_0$, $y' \equiv dy/dz$, thus
\begin{equation}
\label{2.1}
x(z) =  x_0 + x' z
\qquad
y(z) = y_0 + y' z \ .
\end{equation}
The angles $\theta$ and $\phi$ (Fig.~\ref{Fig2}) are given by
\begin{equation}
\label{2.2}
\theta = \arctan \sqrt{ \smash[b]{ {x'}^2 + {y'}^2 } }
\qquad
\phi = \arctan y'/ x' \ .
\end{equation}
Without using the PMTs' times, their positions and charges can be used to obtain an initial estimate of a trajectory
through the method of Least Mean Squares (LMS).
Let $q_i$ be the charge from PMT $i$ at the fixed position $(x_i, y_i, z_i)$,
and let $w_i \equiv q_i^2$ be its weight.\footnote{
  This choice of weights is not mathematically rigorous,
  and there are other possible choices for the weights.
  However, this particular choice is a simple one,
  and found to work better than anything else that was tried, such as $w_i = q_i$ or $w_i = 1$ (unweighted).
  A general aspect of the methods described here is that,
  in places where a difficulty was encountered,
  plausible variations were introduced until there arose a method which overcame the difficulty and functioned as needed,
  with somewhat less regard for rigor and more regard for being able to demonstrate correct functionality.
}
Nine sums are accumulated over $n$ PMT hits in a given event,
\begin{align}
\nonumber
\! S_x &\equiv \! \sum_i \! w_i x_i &
\! S_y &\equiv \! \sum_i \! w_i y_i &
\!\! S_z &\equiv \! \sum_i \! w_i z_i &
\qquad
\\
\nonumber
\! S_{x^2} &\equiv \! \sum_i \! w_i x_i^2 &
\! S_{y^2} &\equiv \! \sum_i \! w_i y_i^2 &
\!\! S_{z^2} &\equiv \! \sum_i \! w_i z_i^2 &
\qquad
\\
\label{2.3}
\! S_{zx} &\equiv \! \sum_i \! w_i z_i x_i &
\! S_{zy} &\equiv \! \sum_i \! w_i z_i y_i &
\!\! S_w    &\equiv \! \sum_i \! w_i \, .\!\! &
\end{align}
Only a small subset of all the PMTs is included in these sums.
This subset is designated as the Use Set, described in detail below.
The slopes and intercepts describing the trajectory are given by
\begin{gather}
\nonumber
 x' = \left( S_z S_x - S_w S_{zx} \right) \big/ \left( S_z^2 - S_w S_{z^2} \right)
\\
\nonumber
 y' = \left( S_z S_y - S_w S_{zy} \right) \big/ \left( S_z^2 - S_w S_{z^2} \right)
\\
\label{2.4}
 x_0 = \left( S_x - x' S_z \right) \big/ S_w \qquad
 y_0 = \left( S_y - y' S_z \right) \big/ S_w \, .
\end{gather}
Though not required by the method, an RPC hit may be incorporated in the sums in Eq.~\eqref{2.3}.
The RPC hit is given a weight equal to the sum of the PMT weights.
This constrains the trajectory to not deviate very much from the RPC hit,
which represents the only truly known point on the trajectory.
Although requiring an RPC hit reduces track-finding efficiency by 40-50\%,
and introduces a bias through reduced angular acceptance,
the trajectories of the surviving tracks are much improved.

\section{PMT Illumination Geometry}
\label{sec3}
\begin{figure}
 \includegraphics[height=4.5cm]{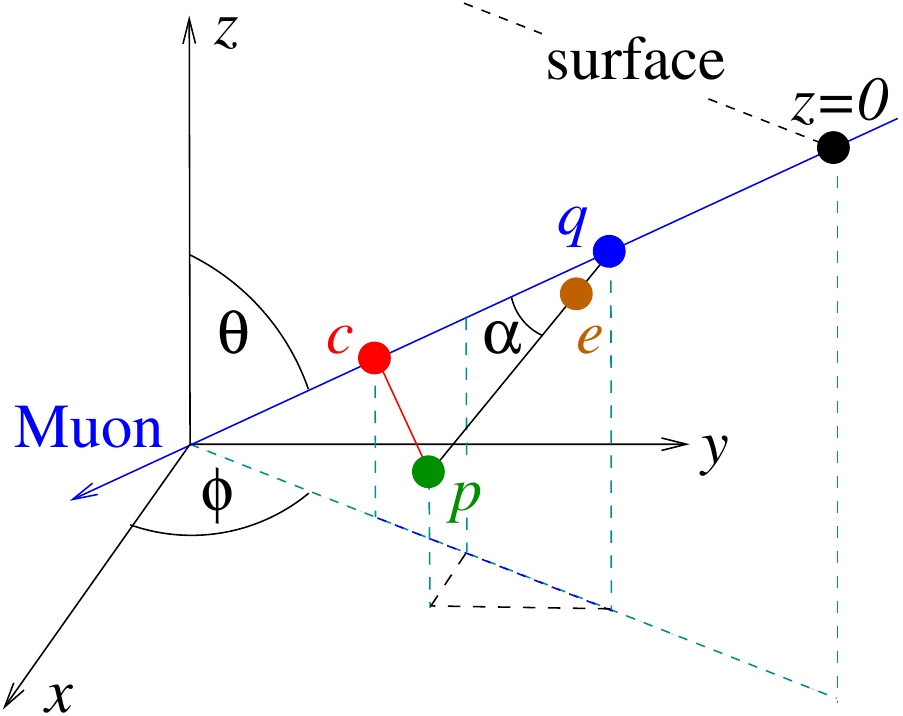}
 \caption{\label{Fig2}
  An hypothesized muon trajectory.
  Point $c$ is the closest approach of the muon to the PMT at point $p$.
  Cherenkov light which illuminates the PMT is emitted from point $q$
  at an angle $\alpha$ with respect to the trajectory.
  The muon shown here enters the pool at a point on the surface, where $z=0$.
  The point $q$ is determined from the PMT's known position, the known angle $\alpha$, and the hypothesized trajectory.
  The point $e$ is determined from those and the PMT's recorded time;
  the effect of the PMT's recorded time is to pull $q$ towards $e$.
  Here, the point $e$ is closer to the PMT than $q$,
  but it is just as common to have $e$ on the other side of $q$.
  The axes shown are shifted from the origin to improve the visualization.
}
 \end{figure}

The following development concerns a single PMT at position ${\mathbf r}_p = ( x_p, y_p, z_p )$,
where the subscripts are used here to distinguish between several useful points shown in Fig.~\ref{Fig2},
rather than between different PMTs as in Eq.~\eqref{2.3}.
By definition, the muon crosses the surface at ${\mathbf r}_0 \equiv (x_0, y_0, 0)$.
For a truly horizontal muon, the algorithm would need some small modifications,
but since horizontal muons are handled in the limit,
with ${\mathbf r}_0$ moving ever further away from the pool for increasingly-horizontal muons,
there is no compelling need for such special treatment.
Besides, if the RPC is used, there will be no tracks anywhere near horizontal.
Considering that the expected resolution for $d_{qp}$ (defined below) is about 50~cm,
which corresponds to the 2~ns PMT timing resolution,
the approximately 10~cm of air between $z=0$ and the actual water surface is of no significant consequence,
especially for muons which cross the surface outside the pool and enter the pool through the side.
Likewise, the finite size (20~cm) of the PMT is not treated because that would introduce substantial complications without
yielding significant benefits.
The trajectory from Eq.~\eqref{2.1} has a point of closest approach to a PMT
${\mathbf r}_c = ( x_c, y_c, z_c )$
given by
\begin{align}
\nonumber
z_c&=\left[ \left( x_p - x_0 \right) x' + \left( y_p - y_0 \right) y' + z_p \right] \cos^2\theta
\\
\label{3.1}
x_c &= x_0 + x' z_c
\qquad
\qquad
y_c = y_0 + y' z_c \, .
\end{align}

Cherenkov light is emitted along a cone characterized by an angle $\alpha$ (from the trajectory),
where $\cos\alpha = 1/n\beta$ and where $c \beta$ is the speed of the muon, with $\beta$ taken as unity.
The index of refraction of ultrapure water $n_W = 1.332986$\cite{pope,fry}, from which
$\alpha = 41.393^\circ$.
\footnote{
While the index of refraction depends on wavelength, only the peak response of the PMTs is available.
From an algorithmic perspective, the only way to handle this would be through a fitting process.
This would greatly complicate and slow the algorithm, with little or no benefit to the results.
}
The point $q$ is on the muon trajectory where Cherenkov light is emitted which hits the PMT at $p$.
At the closest approach, the distance between the muon and PMT is
\begin{equation}
\label{3.2}
d_{cp} = |{\mathbf r}_c - {\mathbf r}_p| \, .
\end{equation}
The distance $d_{qp}$ between points $q$ and $p$ and the distance $d_{cq}$ between points $c$ and $q$ are
\begin{equation}
\label{3.3}
d_{qp} = d_{cp} / \sin\alpha \, , \quad
d_{cq} = d_{cp} \cot\alpha \, .
\end{equation}
The point ${\mathbf r}_q = (x_q, y_q, z_q)$ is given by
\begin{equation}
\nonumber
z_q = z_c + d_{cq} \cos\theta \, , \quad
x_q = x_0 + z_q x' \, , \quad
y_q = y_0 + z_q y' \, .
\end{equation}
The distances from ${\mathbf r}_0$ to ${\mathbf r}_c$ and from 
${\mathbf r}_0$ to ${\mathbf r}_q$ are
\begin{equation}
\label{3.4}
d_{0c} = | {\mathbf r}_0 - {\mathbf r}_c | \, ,
\quad
d_{0q} = d_{0c} - d_{cq} \, .
\end{equation}
Note that $d_{0q}$ is negative if the point $q$ is above the surface,
in which case reflected rather than direct Cherenkov light would be illuminating the PMT.
If $d_{0q}$ is sufficiently negative, the PMT is removed from the sums in Eq.~\eqref{2.3}.
Because of the finite positional resolution, {\em bad-hit tolerances} are applied to all such decisions in order to prevent
the discarding of good hits along with the bad, at the cost of admitting some bad hits into the reconstruction.
These bad-hit tolerances are initially large, but are reduced as the reconstruction is refined, as described below.

Each PMT has a corrected time $t \equiv \tau - t_{0}$,
where $\tau$ is the raw time as given by Eq.~\eqref{1.1},
and where $t_{0}$ is the PMT's timing offset.
Define $T_0$ to be the time at which the muon is just at the surface, relative to the PMTs' corrected times $t_i$.
With the arbitrary value of 1100 in Eq.~\eqref{1.1}, which corresponds to about 1719~ns, $T_0 - 1719$~ns is the
time relative to the trigger that the muon crossed the surface.
%\footnote{
%  Regarding the physical significance of $T_0$:
%  With the arbitrary value of 1100 in Eq.~\eqref{1.1}, which corresponds to about 1719~ns, $T_0 - 1719$~ns is the
%  time relative to the trigger that the muon crossed the surface.
%  Alternatively, defining the muon crossing time as zero, $T_0$ is the time that needs to be subtracted from each PMT's
%  time $t$ to obtain its time relative to the muon crossing time.
%  This is not done because it would complicate what follows with no particular benefit.
%% $T_0$ is observed to have only a weak dependence on the trajectory;
%% the reason for this is not entirely understood, but neither has it been extensively investigated.
%}
$T_0$ is a trajectory parameter, in the same category as $x_0$, $y_0$, etc.
It does not lend itself to direct calculation,
being an event-specific time offset common to all the PMTs seeing light from the same muon.
Rather, $T_0$ is initially set from a constant $T_{00}$, then improved iteratively for each event,
as will be described.
$T_0$ varies from event to event, observed to form a distribution about 30~ns wide.
Because there is a random timing variation between detector subsystems, the IWS and OWS each has its own $T_0$.

The muon travels for a time $t_{0q}$
from the surface ($z_0 \equiv 0$) before arriving at $q$,
where the Cherenkov light is emitted, which light is seen a little later by the PMT at $p$, thus
\begin{equation}
\label{3.5}
% t_{0q} = d_{0q} / c = \left(  d_{0c} - d_{cp} \cot\alpha \right) / c \, .
t_{0q} = d_{0q} / c \, .
\end{equation}
The time for the Cherenkov light to travel from $q$ to $p$,
and the time from when the muon is at the surface to when the light hits the PMT, are
\begin{equation}
\label{3.6}
t_{qp} = d_{qp} n_W / c \qquad t_{0p} = t_{0q} + t_{qp} \, .
\end{equation}
A PMT's corrected time $t$ together with $T_0$ also determines
the time for the muon to travel from the surface to $q$ and the light to reach $p$, thus
\begin{equation}
\label{3.7}
t_{0p}(t) = t - T_0 = \tau - t_0 - T_0 \, .
\end{equation}
The time $t_{0p}(t)$, which depends on the time $t$, is to be distinguished from $t_{0p}$,
which is determined only from the hypothesized trajectory and does {\em not\/} depend on the time.
The difference between these two is
\begin{equation}
\label{3.8}
\delta t = t_{0p}(t) - t_{0p} = \tau - t_0 - T_0 - t_{0p} \, .
\end{equation}
This is a time residual, a measure of how late or early the PMT is from being compatible with the
hypothesized trajectory.
If $T_0$ and the hypothesized trajectory are accurate,
the average of the residuals $\delta t$ over all the PMTs in the sums will be zero.
Define a correction $dT_0$ as the weighted average of $\delta t$ for each of these PMTs,
again using the square charge as the weight.
With $dT_0$, an improved $T_0$ is obtained, thus,
\begin{gather}
\nonumber
dT_0 \equiv \sum_i w_i \delta t_i \Big/ \sum_i w_i
\\ \label{3.9}
T_0 \to T_0 + dT_0\, .
\end{gather}
This will change $t_{0p}(t)$ for all of these PMTs through Eq.~\eqref{3.7},
so the process is repeated until $dT_0$ is less than some fixed minimum $dT_{0\mbox{\tiny min}}$.
Convergence is observed to usually require two iterations,
even with the very small value $dT_{0\mbox{\tiny min}} = 0.001$~ns used here,
and is insensitive to any reasonable starting point $T_{00}$, even zero.\footnote{
  Because this process is {\em never\/} observed to require a third iteration,
  ``convergence'' is perhaps not the correct term to use here.
}
In practice, $T_{00}$ is set to roughly the mean value of $T_0$ as observed from many events.
Because $t_{0p}$ depends on the trajectory, this process of adjusting $T_0$ is repeated whenever the trajectory is adjusted.

The point $e$ is constrained, by definition, to lie somewhere on the line defined by $q$ and $p$ in Fig.~\ref{Fig2}.
The PMT's time $t$ determines where on that line $e$ lies, with $t_{ep}$ the time for light to travel from point $e$ to
point $p$,
\begin{gather}
\nonumber
t_{ep} = t - T_0 - t_{0q} \, , \quad
R_{qe} \equiv t_{ep}/t_{qp} \, , \quad \mbox{and}
\\
\label{3.10}
{\mathbf r}_e = {\mathbf r}_p + R_{qe} \left( {\mathbf r}_q - {\mathbf r}_p \right) \, .
\end{gather}
Using ${\mathbf r}_e$ instead of ${\mathbf r}_p$ in the sums in Eq.~\eqref{2.3}
produces a trajectory improved by making use of the PMTs' times.

Define
\begin{equation}
\label{3.11}
d_{qe} \equiv |\delta t| c / n_W \, .
\end{equation}
This is used in deciding whether a hit is seeing direct light, as will be described shortly.

The chi-squared per degree of freedom $\chi_\nu^2$ is a useful measure of the trajectory goodness.
Although this is not used to fit the trajectory parameters,
it is used to decide whether a trajectory has improved following each iteration,
and is one of several measures used to decide whether a trajectory has been successfully determined.
Let $\delta t_i$ be the difference $\delta t$ from Eq.~\eqref{3.8} for PMT $i$,
and let $\delta_{\mbox{\tiny RPC}}$ be the distance between the track at the RPC and the position of the RPC hit.
Then
\begin{equation}
\label{3.12}
\chi_\nu^2 \equiv \frac{1}{\nu} \left( \frac{ 1 }{ \sigma_t^2 } \sum_i \delta t_i^2
+ \frac{ \delta_{\mbox{\tiny RPC}}^2 }{ \sigma_{\mbox{\tiny RPC}}^2 } \right) \, ,
\end{equation}
where $\sigma_t$ is the PMT time resolution,
$\sigma_{\mbox{\tiny RPC}}$ is the spatial resolution of the RPC,
and where $\nu$ is the number of degrees of freedom.
The RPC term is omitted if the RPC is not used.
% Values of $\sigma_t = 1.5$~TDC~bins = 2.34~ns and $\sigma_{\mbox{\tiny RPC}} = 27$~cm
% are observed to produce a $\chi_\nu^2$ distribution that peaks near unity.
A value of $\sigma_t = 1.2$~TDC~bins = 1.88~ns
is observed to produce a $\chi_\nu^2$ distribution that peaks near unity.
For the RPC, $\sigma_{\mbox{\tiny RPC}} = 10$~cm is used, based on its published resolution \cite{MuonPaper},
but $\chi_\nu^2$ is almost completely insensitive to $\sigma_{\mbox{\tiny RPC}}$ because it is overwhelmed by the PMT data.
For example, from a sample of 25k triggers,
$\sigma_{\mbox{\tiny RPC}}=10$~cm results in 12~158 tracks and $\chi^2_\nu(\mbox{av})=1.405$, whereas
$\sigma_{\mbox{\tiny RPC}}=20$~cm results in 12~163 tracks and $\chi^2_\nu(\mbox{av})=1.402$, and
$\sigma_{\mbox{\tiny RPC}}=5$~cm results in 12~147 tracks and $\chi^2_\nu(\mbox{av})=1.420$.
For a reconstruction using only PMTs in the IWS or OWS, the number of parameters is $N_{\rm par} = 5$,
which includes $x_0, x', y_0, y'$, and $T_0$.
If the reconstruction includes PMTs from both the IWS and OWS, then the number of parameters is $N_{\rm par} = 6$
because a separate $T_0$ is needed for each of the two, as mentioned above.
The number of data $N_{\rm data}$ is the number of PMTs in the sums plus one for the RPC if it is used.
The number of degrees of freedom then is $\nu = N_{\rm data} - N_{\rm par}$.
When not using the PMT times, Eq.~\eqref{3.12} is modified by replacing $\delta_t$ with $d_{cp}$ from Eq.~\eqref{3.2}
and $\sigma_t$ with $\sigma_r = 2.3$~m, with that value chosen to produce a $\chi_\nu^2$ that peaks near unity.
In this case $N_{\rm par} = 4$,
since there are no $T_0$ parameters.

As a further check of track veracity, the residuals in Eq.~\eqref{3.8} can be combined to obtain
an estimate of the muon speed $\beta_\mu$.
Define these sums, in analogy with Eq.~\eqref{2.3},
\begin{gather}
\nonumber
S_d     \equiv \! \sum_i \!\!  w_i t_{0qi}             \quad
S_{td}  \equiv \! \sum_i \!\!  w_i t_{0qpi}(t_i) t_{0qi} \quad
S_w     \equiv \! \sum_i \!\!  w_i                     \\
\nonumber
S_t     \equiv \! \sum_i \!\!  w_i t_{0qpi}(t_i)         \quad
S_{d^2} \equiv \! \sum_i \!\!  w_i t_{0qi}^2           \quad
S_{t^2} \equiv \! \sum_i \!\!  w_i t_{0qpi}^2(t_i)       \\
\nonumber
t_{0qpi}(t_i) \equiv t_i - T_0 - t_{qpi} \qquad 
w_i \equiv 1 \, .
\end{gather}
where $t_{0qi}$ and $t_{qpi}$ are $t_{0q}$ from Eq.~\eqref{3.5} and $t_{qp}$ from Eq.~\eqref{3.6} for PMT $i$,
both of which depend only on the hypothesized trajectory.
The resulting $t_{0qpi}(t_i)$ depends on the PMT offset-corrected time $t_i$.
It might be more apparent that this renders a speed if $d_{0qi}$ were used instead of $t_{0qi}$,
but this would need to be divided by $c$ to keep the units of the sums' products the same in what follows,
and this is just $t_{0q} = d_{0q}/c$ from Eq.~\eqref{3.5}.
Then, in analogy with a slope from Eq.~\eqref{2.4},
\begin{gather}
\nonumber
\beta_{\mu 0} \equiv \left( S_t S_d - S_w S_{td} \right) \big/ \left( S_d^2 - S_w S_{d^2} \right) \\
\nonumber
\beta_{\mu 1}^{-1} \equiv \left( S_t S_d - S_w S_{td} \right) \big/ \left( S_t^2 - S_w S_{t^2} \right) \\
\label{Eq:MuV}
\beta_\mu = \sgn \left( S_{td} \right) \sqrt{ \beta_{\mu 0} \beta_{\mu 1} } \, .
\end{gather}
This last is the geometric mean of velocities obtained from the slope $ds/dt$ and the inverse of the slope $dt/ds$,
which is equivalent to swapping abscissa $t$ and ordinate $s$,
where $s$ is the distance traveled by the muon divided by $c$.
In the case of poorly correlated points, the slope $ds/dt$ obtained from LMS pulls towards zero,
whereas the geometric mean pulls towards unity, which is more in line with expectations.
(This is not done for the track slopes from Eq.~\eqref{2.4}, where there are no firm expectations,
but this causes less-well-determined tracks to be pulled towards the vertical.)
This forms a distribution with mean $\beta_\mu = 1$ and an RMS of about 0.2.
If the geometric mean were not employed, the mean would be about 20\% below 1,
pulled towards zero by tracks suffering from
closely-spaced clusters of points $r_{pi}$ with large residuals $d_{qei}$.
Because $T_0$ is different for each detector,  $\beta_\mu$ is determined separately for each detector, then averaged.
More than anything, this is a convenient way of expressing the residuals, and serves as an additional sanity check which
catches some bad tracks which otherwise have a reasonable $\chi_\nu^2$.

As shown, the sums in Eq.~\eqref{Eq:MuV} are unweighted.
To ameliorate the problem of nonsense speeds due to the clustering described above,
a weight which strongly disfavors hits with $d_{qei} \gg d_{qe0}$ can be used, such as
$w_i \equiv d_{qe0}^2 / \left( d_{qe0}^2 + d_{qei}^2 \right)$,
where $d_{qe0}$ is somewhat arbitrarily set to 0.2~m.
However, such an improved speed is less useful in identifying poorly reconstructed tracks.
\section{The Algorithm}
\label{sec4}
The algorithm treats only the major obstructions, i.e., the pool walls, floors of the IWS and OWS,
the surface of the pool, and the ADs.
Given the expected reconstruction resolution of about 50~cm, it is not sensible to account for 
the AD stands, various objects on the ADs, cables, plumbing, and other such items,
especially considering the cost in CPU cycles for doing so,
not to mention the unrealistic requirement of knowing exactly where all such objects are.
While the algorithm can find multiple tracks, this feature is usually disabled since its reliability has not been verified
with Monte Carlo studies.
The track-finding process starts by collecting the PMT data from each detector in an event into a {\em hit list}.
If there are fewer than $N_{\rm useMin}$ hits in the hit list, no attempt is made to find a track.
$N_{\rm useMin}=5$ is the minimum for the degrees of freedom to be greater than zero,
but this admits some poor, even accidental tracks, so in practice $N_{\rm useMin}=8$ is taken.

An RPC hit is stored separately from the hit list.
Events with multiple RPC hits are a small fraction of the total, so no effort is expended in looping over RPC hits,
i.e., looking for the best match to the PMT data.
Rather, events with multiple RPC hits are simply rejected if the RPC is used.
If the RPC is not being used, $N_{\rm useMin}$ is increased by one.

\begin{figure}
 \centering
 \includegraphics[width=0.5\textwidth]{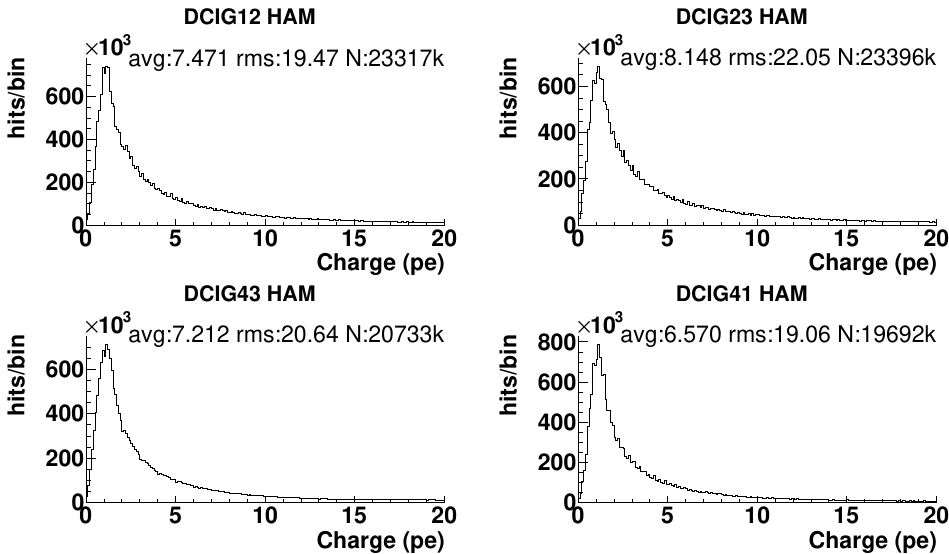}
 \includegraphics[width=0.4\textwidth]{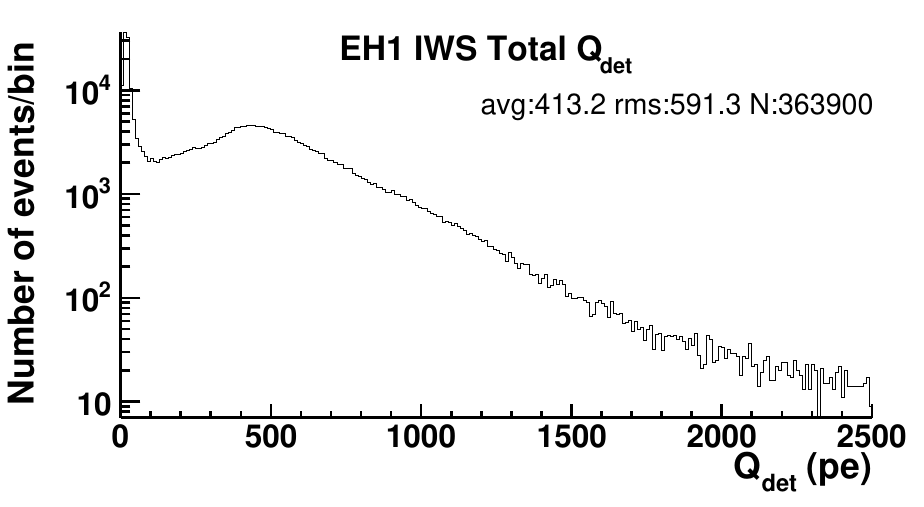}
 \caption{\label{Fig3}
    {\bf Top} Charge distributions of four typical IWS PMTs, from 45M IWS triggers.
    These have very long tails (not shown), extending to about 1000~pe, though a few
    counts are as high as about 2000~pe.
    {\bf Bottom} Whole-detector charge distributions for the IWS PMTs from the same data set.
    The OWS distribution is similar. This shows all IWS data, without requiring reconstructed tracks.
 }
 \end{figure}

The hit with the largest charge is identified as hit$_0$.
Up to $N_{\rm hitMax}$ additional hits with the largest charges are then selected
for inclusion in the initial Use Set using the following criteria:
\begin{enumerate}
\item A hit must be on the same wall as hit$_0$, or on an immediately adjoining wall,
which includes the floor.
Because the initial set is preferred to not include hits from all walls,
and since all walls adjoin the floor,
only hits from the floor are selected for the initial set if hit$_0$ is on the floor.
\item A hit must have charge greater than the somewhat arbitrary (but reasonable -- see Fig.~\ref{Fig3}, top) value of 2~pe.
\end{enumerate}
Studies have shown that $N_{\rm hitMax} = 40$ is close to optimal.
Likewise, the minimum charge of 2~pe has been determined by studies.
Setting this threshold much lower allows too many ``bad hits'' (those with reflected vs. direct light) to be included,
and reduces the reconstruction efficiency.

Only hits from the Use Set are included in the nine sums in Eq.~\eqref{2.3},
which are accumulated and applied to Eq.~\eqref{2.4} to obtain an initial, estimated trajectory.
The track's $T_0$ values for the IWS and OWS are obtained, and $\chi_\nu^2$ is calculated.
Initial bad-hit tolerances are set, allowing for a poor initial trajectory.

The calculations described in \S \ref{sec3} are performed for each PMT in the hit list,
regardless of previous membership in the Use Set,
with the following criteria applied to determine if it is expected to be seeing direct light.
If so, it is included in the new Use Set.
\begin{enumerate}
\item The distance (or residual) $d_{qe}$ must be smaller than the current bad-hit tolerance $d_{\rm tol}$.
\item The points ${\mathbf r}_q$ and ${\mathbf r}_p$ (Fig.~\ref{Fig2}) must be
on the same side of every surface intersected by the line defined by the segment $qp$
(walls, floors, water surface, and the top, bottom and sides of the ADs),
within the bad-hit tolerance $s_{\rm tol}$ (distinct from $d_{\rm tol}$) of a surface's edge.
\item The dot product of ${\mathbf r}_{qp}$ and the direction vector of the PMT must be less than
the bad-hit tolerance $r_{\rm tol}$,
indicating that the PMT is facing in a suitable direction to see direct light.
This dot product is -1 for head-on light, whereas light coming from behind a PMT has a value of +1.
Initially $r_{\rm tol}=0.866$, corresponding to 60$^\circ$ in the {\em wrong} direction.
\end{enumerate}

The track is next iteratively refined.
In each iteration, the nine sums in Eq.~\eqref{2.3} are accumulated using
the points ${\mathbf r}_e$ from Eq.~\eqref{3.10} instead of the fixed PMT positions ${\mathbf r}_p$.
Equation~\eqref{2.4} is applied to those accumulated sums to obtain the refined trajectory.
With the new trajectory, the calculations described in \S \ref{sec3} are performed again,
and membership in the Use Set redetermined.
The track's $T_0$ values for the IWS and OWS are adjusted, and a new $\chi_\nu^2$ is calculated.
If the track has too few hits in the Use Set, or
if a track's charge summed over all its hits is less than $Q_{\rm sumMin}=300$~pe (determined by studies),
then the track is rejected.
If the bad-hit tolerance $d_{\rm tol}$ is greater than 1~m,
it is reduced so as to exclude as many as six hits with the largest values of $d_{qe}$ (but none with $d_{qe}<1$~m).
This tolerance is never reduced below 1~m, since that corresponds roughly to twice the PMT time-resolution.
If $s_{\rm tol}$ is greater than 0~m, it is similarly reduced,
as is $r_{\rm tol}$ if it is greater than zero (unitless).
The choice of six for the maximum number of hits to exclude in one iteration is based on studies which sought to increase
speed without compromising track-finding veracity.
If there are fewer than ten hits on a track, no more than one hit is excluded in each iteration.
It is observed that if the bad-hit tolerances are reduced too rapidly,
then some good hits are removed along with truly bad ones, before those good hits have had an opportunity to improve the track.

The track is considered fully refined when all of these conditions are satisfied:
\begin{enumerate}
\item $d_{\rm tol}$ has been reduced to 1~m.
\item $s_{\rm tol}$ has been reduced to 0~m.
\item $r_{\rm tol}$ has been reduced to 0.
\item The surface crossing point changes by less than 0.01~cm, as determined by $x_0$ and $y_0$.
\end{enumerate}
Reconstruction terminates under any of the following conditions, some of which result in rejection of the track:
\begin{enumerate}
\item If the track is fully refined.
\item If the number of hits in the Use Set falls below $N_{\rm useMin}$, the track is rejected.
\item If $\chi_\nu^2$ has gotten worse following an iteration, an attempt is made to discard additional hits.
If two such attempts fail to improve $\chi_\nu^2$, the track is accepted as-is if $\chi_\nu^2 < 5$,
otherwise the track is rejected.
\item If the number of iterations exceeds 2000, the track is accepted as-is.
\end{enumerate}
% When the track is fully refined, a final acceptance check is performed to ensure that
% the confidence level of the track, calculated from $\chi^2_\nu$ and $\nu$, is greater than 0.01.
% If not, the track is rejected.
% The choice of the cut at 0.01 was determined from an uncut confidence level distribution, which has
% an increasing number of entries for values below 0.1, and more than twice as many below 0.01 as between 0.01 and 0.02.
% a large number of entries below that value.
% Finally, if the estimated speed of the muon is not positive, the track is rejected.
When reconstruction successfully terminates, a check is made on the estimated muon speed.
If not positive, the track is rejected.

When a track is rejected the attempt is renewed, but starting on a different wall than that of
the (previous) initially selected hit.
This process continues until either a track has been found, or all of the walls have been tried and no track found.

Earth muons (upward moving, where a high energy neutrino has passed through the Earth and created a muon just below Daya Bay)
may be treated by taking $d_{cq}$ as negative in Eq.~\eqref{3.3}
and reversing the sign in the relation between time and distance along the muon path, i.e. in Eq.~\eqref{3.5}.
When searching for Earth muons, a more stringent though somewhat arbitrary value of $N_{\rm useMin}=14$ is used.
Earth muons have an additional constraint imposed on them, since these are so rare and subject to accidental identification.
This constraint consists of stepping through each PMT in the detectors being used, without regard to whether it is present in
the readout, and calculating whether it is expected to have been seen and included in the Use Set.
These are counted, separately for each detector being used, and compared with the corresponding number of hits in the Use Set.
If hits in the Use Set are fewer than half as many as expected for each detector, the track is rejected as an Earth muon.
This constraint has little effect in the search for cosmic muons, and is not used there.
The muon velocity $\beta_\mu$ is required to be negative, indicating that it is upward-going.
Finally, if a prospective Earth muon is identified, the event is searched for a cosmic muon.
If one is found, the Earth muon is rejected.
As of this writing, no Earth muons have been identified with this method.

\section{Distributions}
\label{sec5}
\begin{figure}
   \centering
   \includegraphics[width=0.5\textwidth]{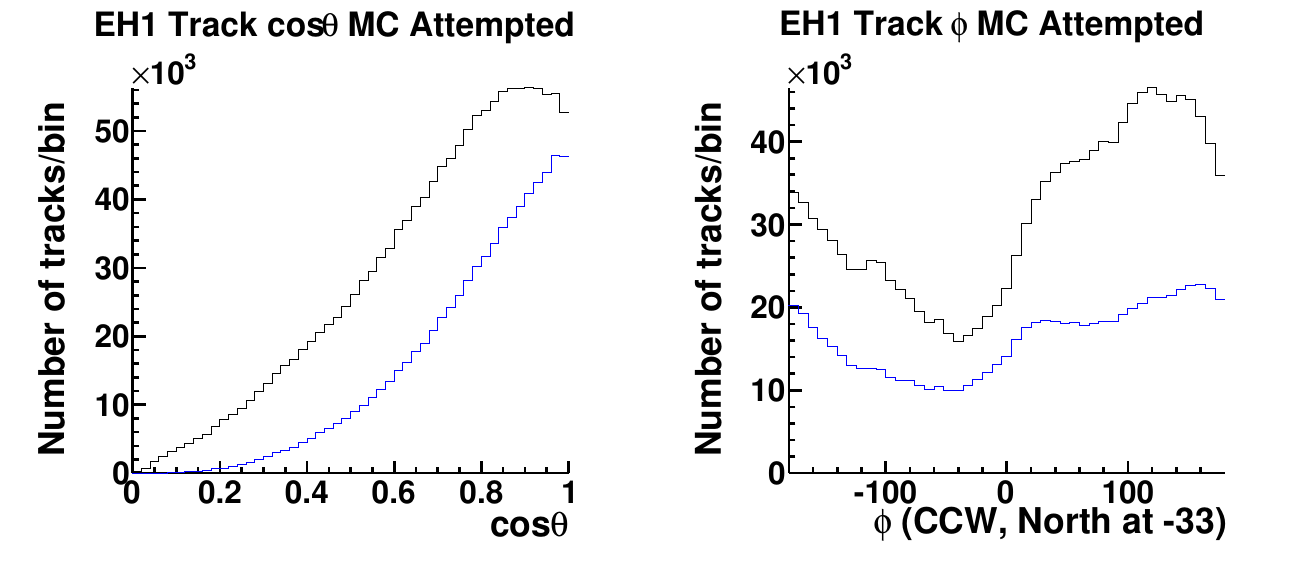}
   \includegraphics[width=0.5\textwidth]{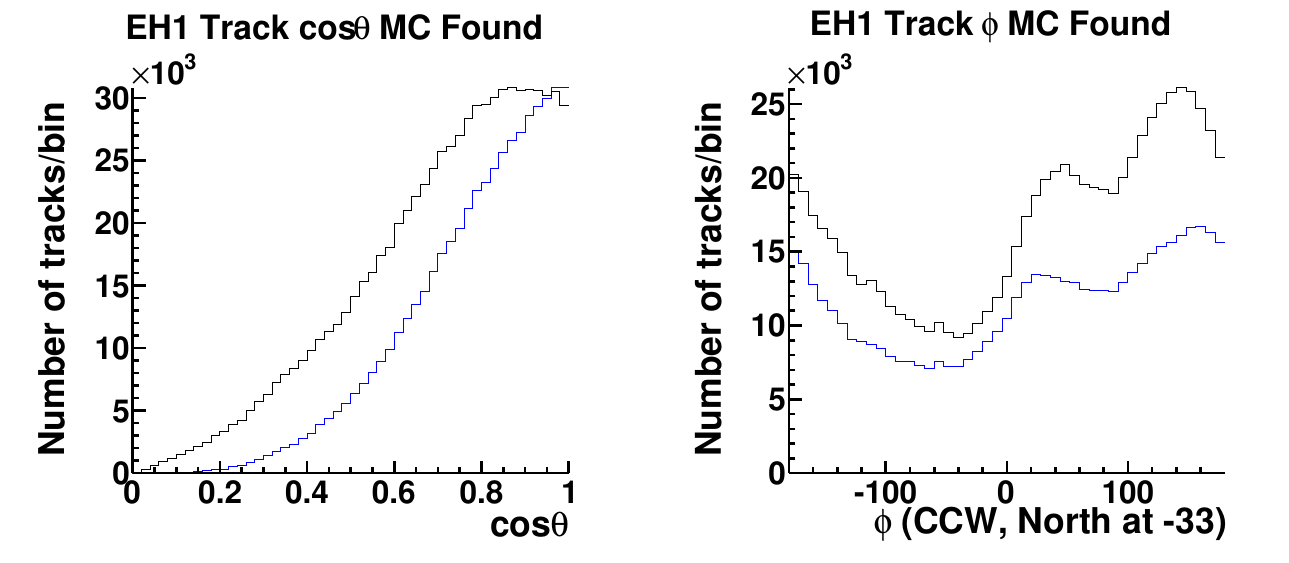}
   \includegraphics[width=0.5\textwidth]{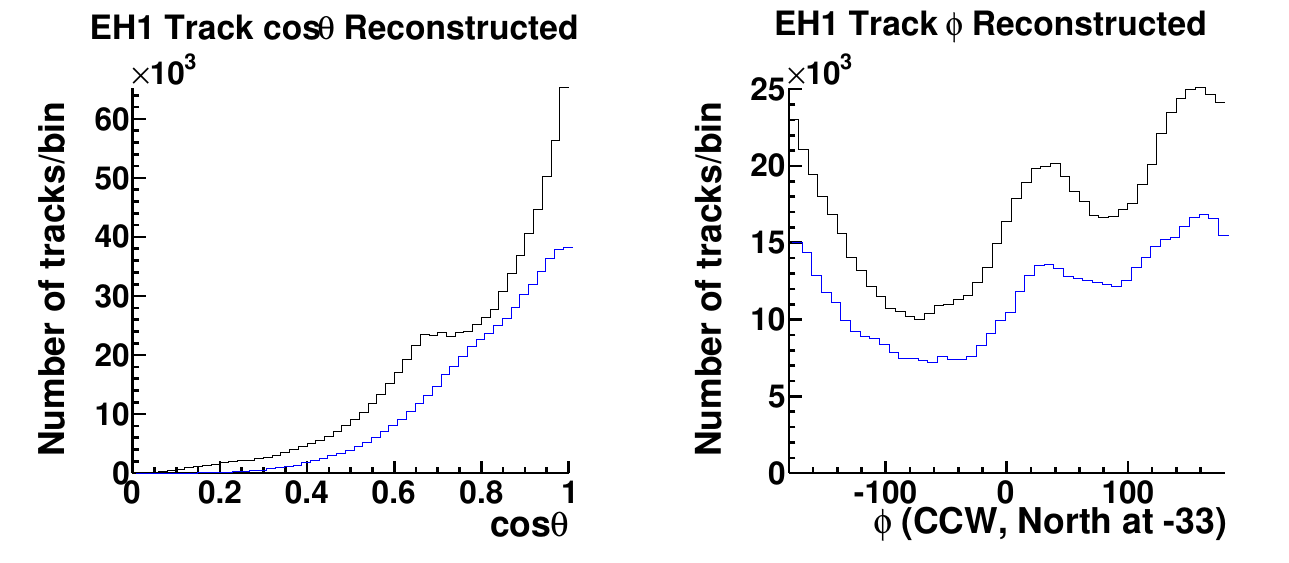}
   \includegraphics[width=0.5\textwidth]{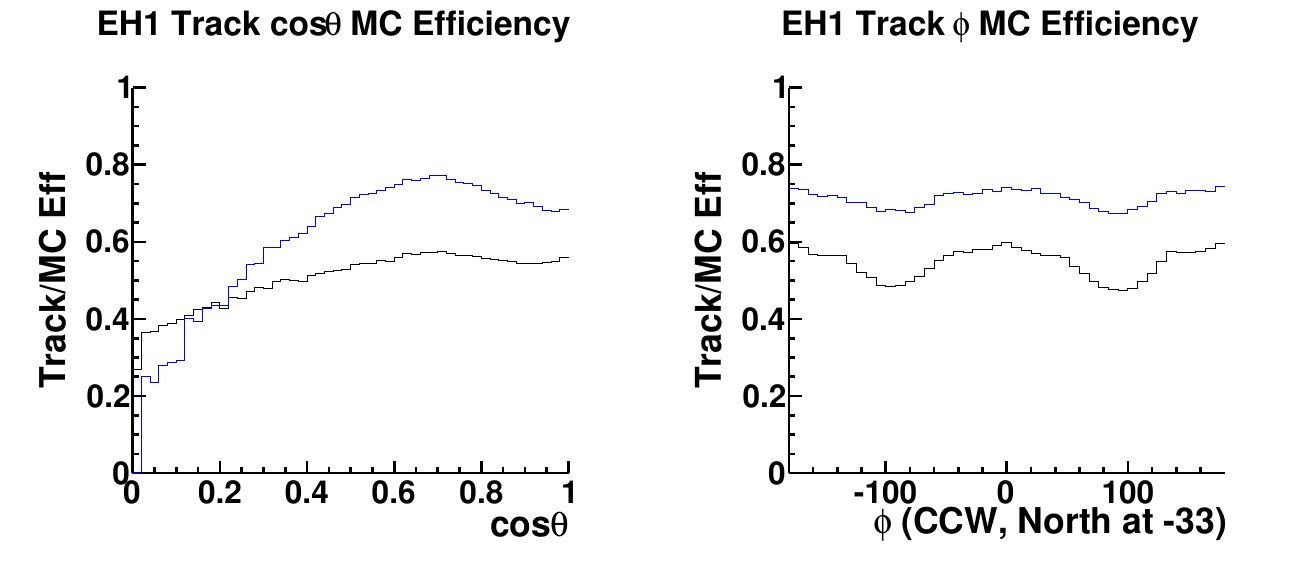}
   \includegraphics[width=0.5\textwidth]{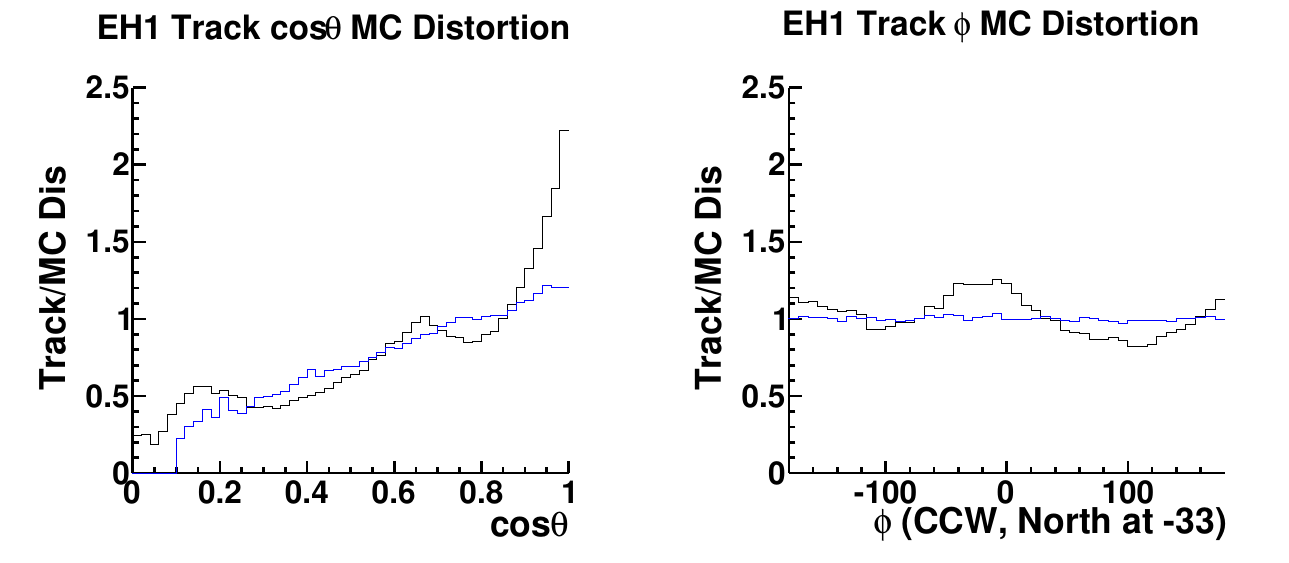}
 \caption{\label{Fig4}
   Comparison of MC truth and the corresponding reconstructed
   track $\theta$ and $\phi$ angular distributions.
   Black: RPC not used.  Blue: One RPC hit required.\\
   {\bf Attempted:} MC truth for simulated muons with hits in one or both of the IWS and OWS, i.e.,
   only those where reconstruction is attempted.
   The $\phi$ distribution represents the simulated effect of the mountain at Daya Bay.\\
   {\bf Found:} MC truth for successfully reconstructed tracks.\\
   {\bf Reconstructed:} Reconstructed values for successfully reconstructed tracks
   (same number of entries as Found).\\
   {\bf Efficiency:} Bin-by-bin ratios of Found to Attempted.\\
   {\bf Distortion:} Bin-by-bin ratios of Reconstructed to Found.
 }
\end{figure}
Figure~\ref{Fig4} shows the angular distributions of tracks reconstructed from Monte Carlo (MC) data for EH1,
the only hall for which such data is available.
This data set comprises 1024k IWS, 1358k OWS, and 1227k RPC triggers, from which 405k tracks were reconstructed when the RPC
was required, and 763k reconstructed when the RPC was not required.
The $\theta$ efficiency plot reveals a detector bias which increasingly disfavors increasingly horizontal tracks
as well as the most vertical tracks.
This is primarily caused by two effects:
\begin{enumerate}
\item Fewer floor PMTs seeing direct Cherenkov light as tracks become more horizontal.
\item Muons which pass nearly through the center of an AD (either horizontal or vertical)
will have a very short path through the water,
and only a few PMTs will be directly illuminated by the Cherenkov light cone.
\end{enumerate}
Neither of these effects has any impact on the function of the muon veto,
since in all cases there is plenty of {\em reflected} light.
The impact is strictly on reconstruction, because of the reduced amount of {\em direct} light illuminating PMTs.
The second effect is most likely also responsible for the small apparent loss for $\phi\sim\pm 90^\circ$,
as seen in the $\phi$ efficiency plot of Fig.~\ref{Fig4}.
Some fraction of such muons will pass nearly through the center of an AD,
losing 5~m or more of path-length through water, with a corresponding loss in reconstruction efficiency.
Tracks with $\phi\approx 0^\circ$ or $\phi\approx 180^\circ$ would have to pass through two ADs, which is a smaller fraction
of all tracks.
The effect for these angles would therefore be correspondingly smaller, and is not obviously visible.
The distortion plots at the bottom of Fig.~\ref{Fig4} shows that $\theta$ is systematically
reconstructed more vertical than MC truth, especially when the RPC is not used,
while there is no such systematic error in $\phi$ when the RPC is used.

\begin{figure}
 \centering
 \includegraphics[width=0.5\textwidth]{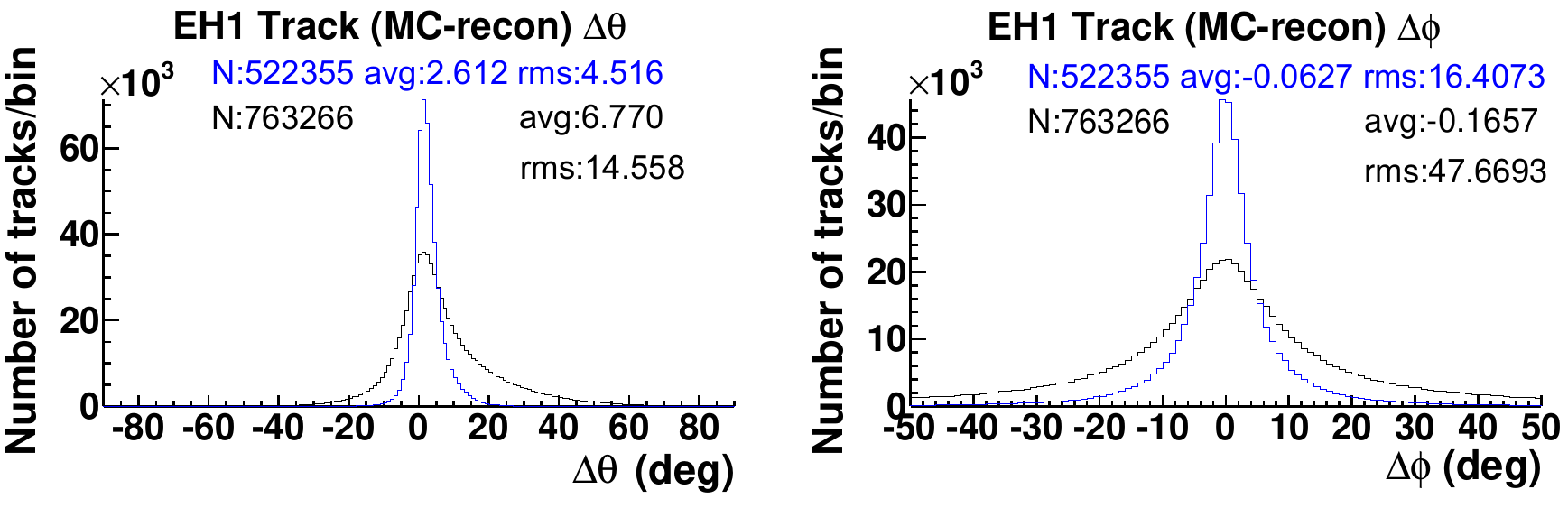}
 \includegraphics[width=0.5\textwidth]{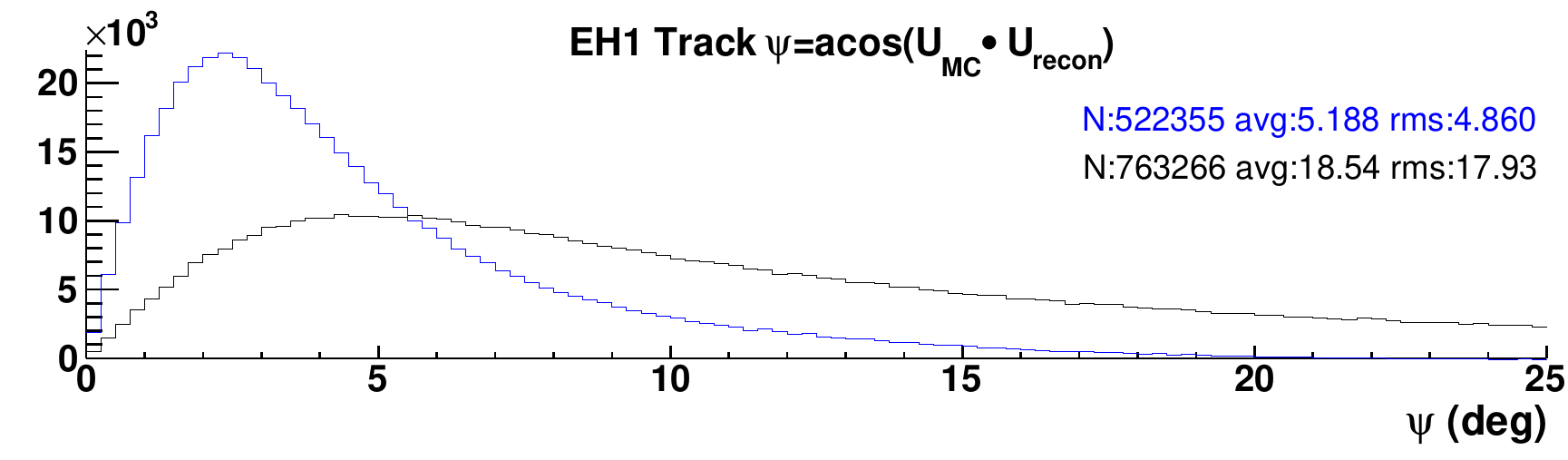}
 \includegraphics[width=0.5\textwidth]{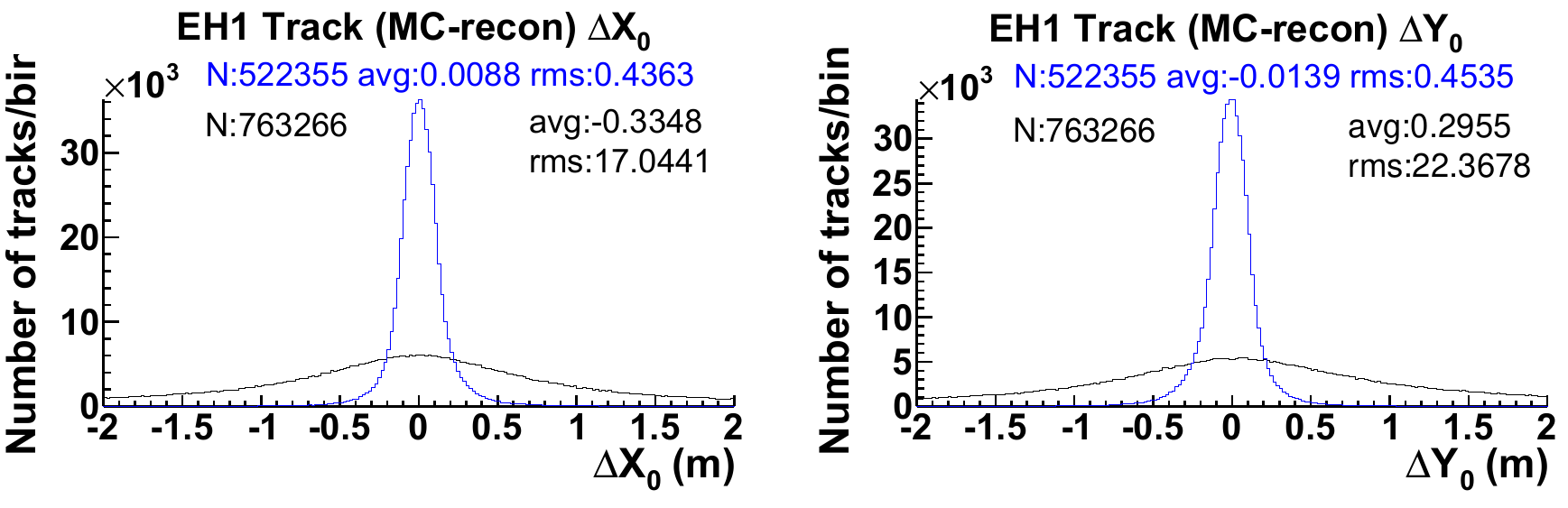}
 \caption{\label{Fig5}
    Differences between MC truth and reconstructions.
    Black: RPC not used.  Blue: One RPC hit required.\\
    {\bf Top}: Separate $\theta$ and $\phi$ directional differences.\\
    {\bf Middle}: Overall directional difference (see text).\\
    {\bf Bottom}: Surface-crossing point $x$ and $y$ differences.
 }
 \end{figure}

Figure~\ref{Fig5} shows the reconstruction errors as differences between MC truth and reconstructed values.
The vertical distortion can plainly be seen in the top plot,
where the average difference shows $\theta$ to be almost $3^\circ$ less vertical in reconstructions than MC truth.
However, the most-vertical tracks are pulled oppositely, with the reconstructed tracks being more vertical than MC truth,
especially when the RPC is not used,
as evidenced by the $\cos\theta$ distortion plot of Fig.~\ref{Fig4}.
It is evident that there is no systematic shift in $\phi$.
The overall directional error is computed as the angle $\arccos({\bf U}_{\rm MC} \cdot {\bf U}_{\rm recon})$,
where ${\bf U}_{\rm MC}$ is a unit vector giving the direction of the track from MC truth information,
and where ${\bf U}_{\rm recon}$ is a unit vector giving the direction of the reconstructed track.
The directional error is given by the average of this angle, and is about $5^\circ$.
Most reconstructions have a directional error of less than $3^\circ$, but the distribution has a long tail.
The positional error in the surface crossing is given by the RMS of the $x$ and $y$ difference distributions,
and is about 45~cm.
Those are the values obtained when the RPC is used.
The situation is considerably worse when the RPC is not used.
 \begin{figure}
 \centering
 \includegraphics[width=0.5\textwidth]{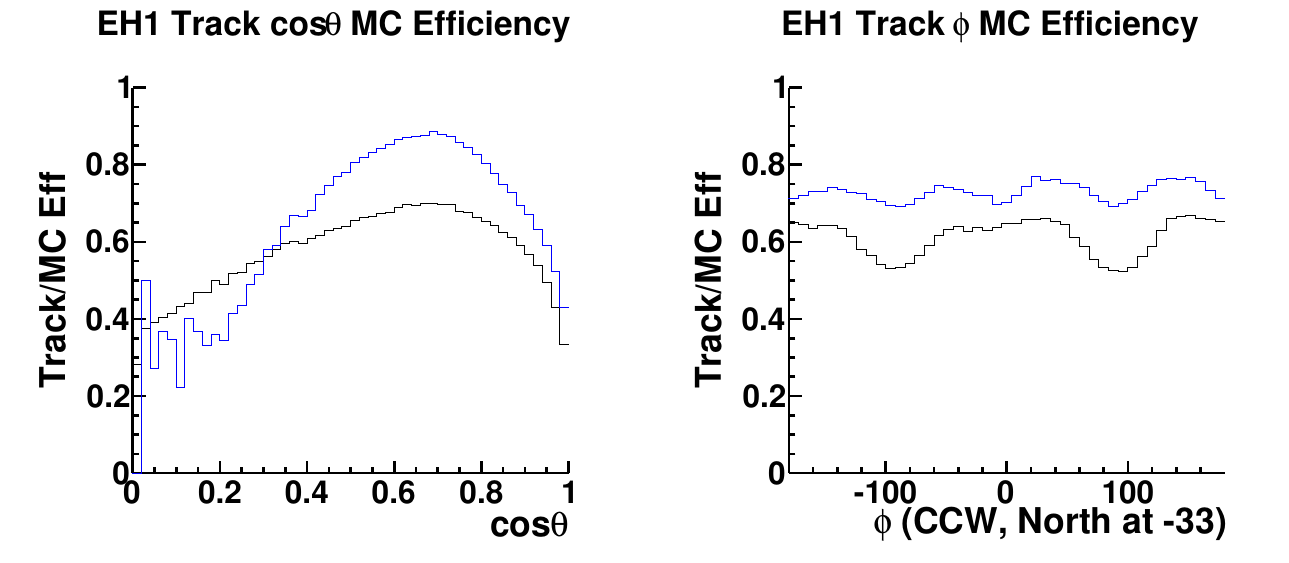}
 \caption{\label{Fig6}
    The same as the efficiency plot in Fig. \ref{Fig4},
    but requiring hits in the Use Set for both IWS and OWS in the reconstructed tracks.
    Black: RPC not used.  Blue: One RPC hit required.
 }
 \end{figure}
 \begin{figure}
 \centering
 \includegraphics[width=0.5\textwidth]{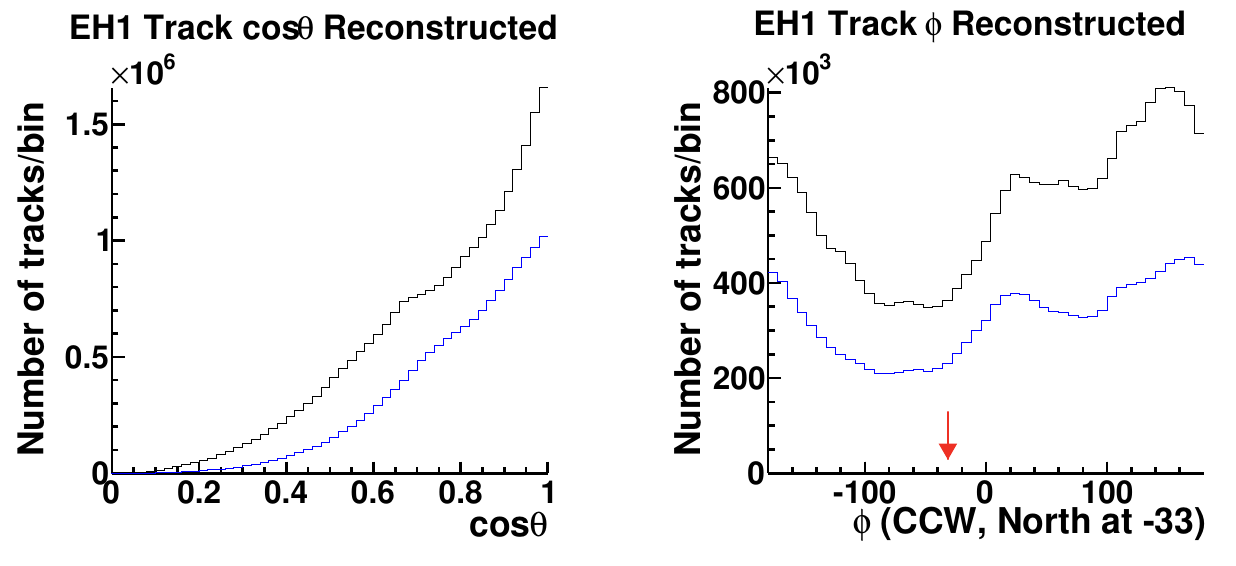}
 \includegraphics[width=0.5\textwidth]{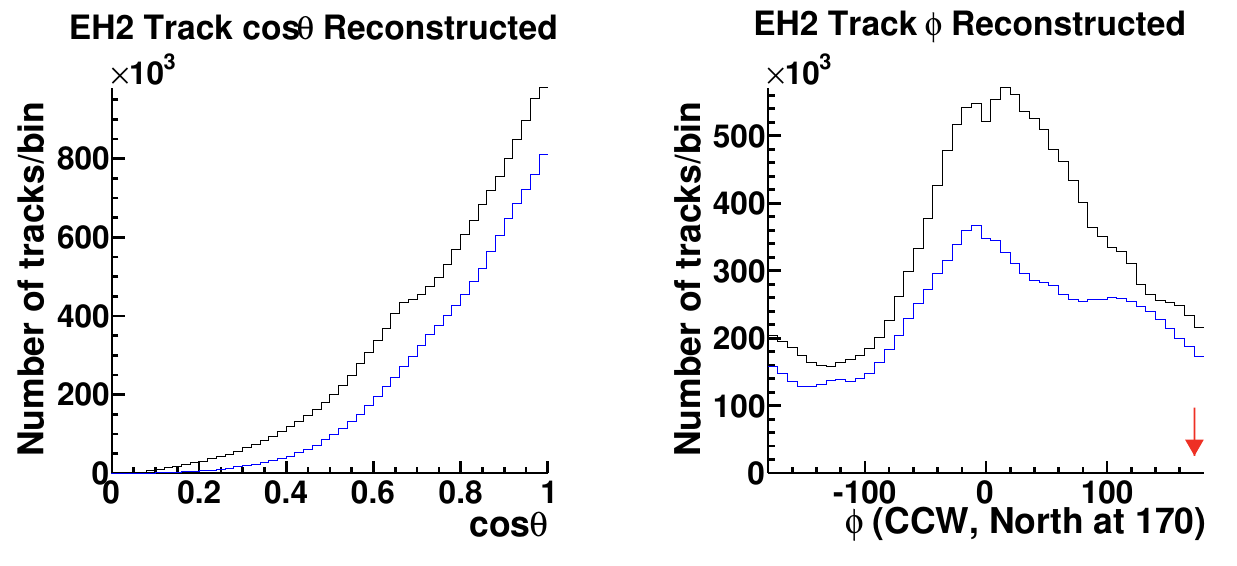}
 \includegraphics[width=0.5\textwidth]{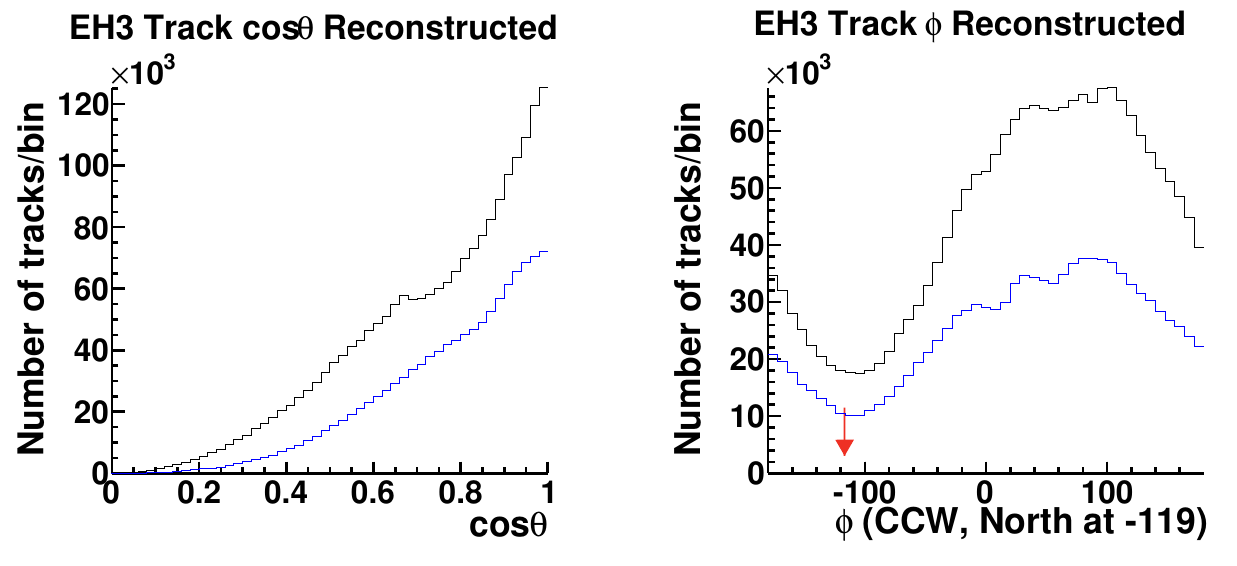}
 \caption{\label{Fig7}
    Reconstructed track $\theta$ and $\phi$ for the three halls, from real data.
    The arrows indicate North.
    Black: RPC not used.  Blue: One RPC hit required.
    Total reconstructions without the RPC: 25M for EH1, 15M for EH2, and 2M for EH3.
    These data comprise
    45M IWS, 69M OWS, and 48M RPC triggers for EH1;
    32M IWS, 40M OWS, and 22M RPC triggers for EH2; and
     8M IWS, 16M OWS, and 12M RPC triggers for EH3.
 }
 \end{figure}
 \begin{figure*}
 \centering
 \includegraphics[width=\textwidth]{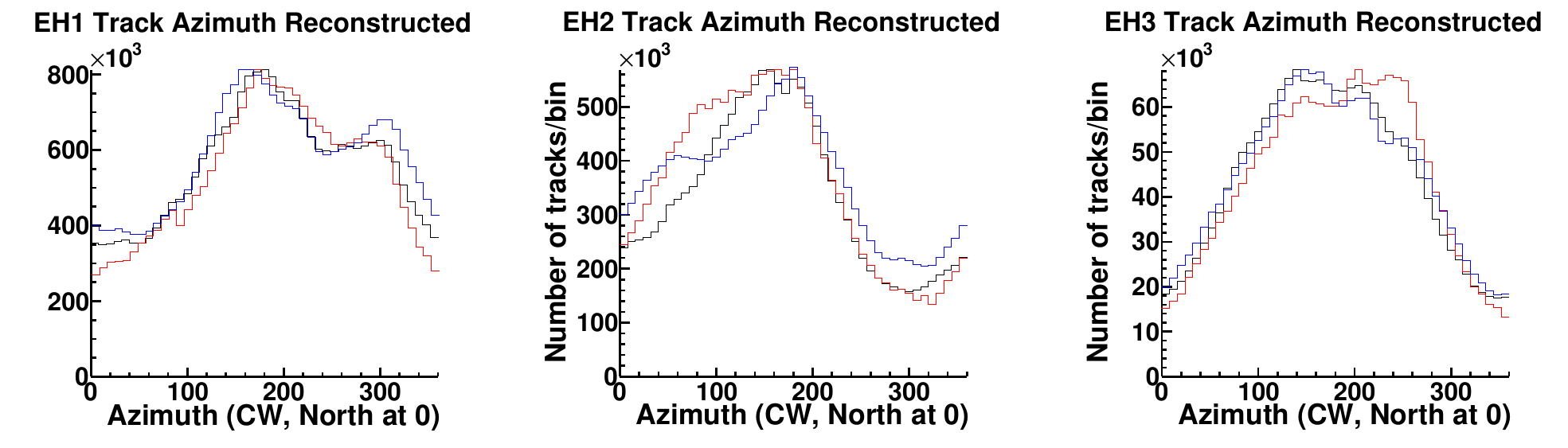}
 \caption{\label{Fig8}
    Reconstructed track azimuth distributions for the three halls, using the same data as in Fig.~\ref{Fig7}.
    Black: RPC not used.  Blue: One RPC hit required, vertically rescaled to match the no-RPC plot.
    Red: Simulated distributions, vertically scaled and binned to match the no-RPC plot.
    Each simulation comprises one million muons as described in \cite{MuonPaper}.
 }
 \end{figure*}

Requiring hits in both the IWS and OWS introduces a peculiar bias, as can be seen in Fig.~\ref{Fig6}.
This bias disfavors tracks running parallel to the pool walls,
causing losses for tracks with $\cos\theta\approx1$ or $\phi\approx0^\circ$, $\pm 90^\circ$, and $180^\circ$:
A track nearly parallel to a wall is less likely to be seen by PMTs in both the IWS and OWS.
To avoid this bias, hits from both IWS and OWS are not generally required for track reconstruction.

Figure~\ref{Fig7} shows the reconstructed angular distributions for the three halls from
about four days worth of data taken in 2012.
Because of the mountain to the North, more muons arrive from the South, as can easily be seen with the aid of the
arrows indicating North.
Figure~\ref{Fig8} shows the reconstructed azimuth distributions for the three halls,
along with the expected (simulated) distributions from Fig.~11 of \cite{MuonPaper}.
The reconstructed distributions match the simulated distributions fairly well.

\section{Obtaining the PMT Time Offsets from Muons}
\label{sec6}
For the purpose of determining time offsets, a special, high-quality subset of the Use Set is designated as the Select Set.
Membership in the Select Set is granted for each hit with
$q > Q_{\rm hitMin}$ and $d_{cp} < d_{cp{\rm Max}}$, where $d_{cp}$ is the closest approach of the muon to a particular PMT
(Fig.~\ref{Fig2}).
Studies have shown that nearly optimal values are $Q_{\rm hitMin} = 50$~pe,
$d_{cp{\rm Max}} = 1$~m for the OWS (corresponding to the space between the walls or floor and the inner Tyvek), and
$d_{cp{\rm Max}} = 5$~m for the IWS (which has more open space than the OWS).
There must be at least one hit in the Select Set from each detector being used,
and a total for three such hits altogether in EH1 and EH2.
Because of the lower statistics in EH3, the latter requirement is relaxed there to require only two such hits altogether.
The benefits of requiring both IWS and OWS hits outweigh the cost of the previously mentioned bias it introduces,
which is irrelevant for the purpose of determining offsets.
% A minimal set of information describing the hits with Use=true
% is appended to a special {\em event list}.

So far, the development has been concerned with single events, summing over PMTs in one event at a time.
What follows concerns a large collection of events, summing over events as well as PMTs.
When obtaining offsets, no more than one track is reconstructed in a given event.
For PMT $i$ and event $j$,
$\tau_{ij}$ is the raw PMT time,
$t_{ij} \equiv \tau_{ij} - t_{0i}$ is the corrected time of PMT $i$,
$t_{0i}$ is the PMT's timing offset,
$t_{0pij}$ is the geometry-derived expected time defined in Eq.~\eqref{3.6},
and $\delta t_{ij}$ is the difference, or residual, defined in Eq.~\eqref{3.8} and which contains the PMTs' time information.
$T_{0j}$ is $T_0$ as obtained iteratively for each event $j$ through Eq.~\eqref{3.9}, with separate values for the IWS and OWS.
Because of these detector-specific values for $T_0$, what follows is performed independently on the IWS and OWS,
regardless of whether tracks were found using the IWS and OWS separately or together.
With the number of PMTs $N_{\mbox{\tiny PMT}}$,
define $N_{\mbox{\tiny PMT}}$ sets of events $\{S_i\}$,
which are those events that have a hit in PMT $i$ and which is a member of the Select Set.
Define a quantity to be minimized, summed over all PMTs $i$ and events $j \in \{S_i\}$, thus
\begin{equation}
\label{5.1}
\chi^2 =  \sum_i \sum_{j\in\{S_i\}} \delta t_{ij}^2 \, .
\end{equation}
Note that this double sum, which is over PMTs and selected events for each PMT,
is {\em not\/} weighted, which is to say that all hits in the Select Set in all events have equal weight.
Weighting this sum, e.g. with the square charge of a hit,
is {\em observed\/} to degrade the algorithm, in contrast to the situation with $T_{0j}$, where
the sum is over PMTs in a single event, and where {\em not \/} weighting degrades the algorithm for obtaining $T_{0j}$.
Evidently, the charge is effective in weighting PMTs in a given event, but not across events.
Expanding this to show the hits' times,
\begin{align}
% \nonumber
% \chi^2 &=
% \sum_i \sum_{j\in\{S_i\}} \left( \tau_{ij} - t_{0i} - T_{0j} - t_{0pij}\right)^2
% \\
\nonumber
\chi^2 &=
\sum_i \sum_{j\in\{S_i\}} \left( t_{0i} - U_{ij}\right)^2 \, , \, \mbox{with}
\\
\label{5.2}
U_{ij} &\equiv \tau_{ij} - T_{0j} - t_{0pij} \, .
\end{align}
$U_{ij}$ varies from event to event, but $t_{0i}$ does not.
Clearly, $d\tau_{ij}/dt_{0i} = 0$, $dt_{0pij}/dt_{0i} = 0$, and $dt_{0k}/dt_{0i} = \delta_{ki}$ (the Kronecker delta).
While $T_{0j}$ (and therefore $U_{ij}$) depends on the time offsets $t_{0i}$ through Eqs.~\eqref{3.7} - \eqref{3.9},
it turns out that this otherwise highly complicating dependence can be ignored to first order, i.e.,
take $dT_{0j}/dt_{0i} = 0 \Rightarrow dU_{ij}/dt_{0i}=0$,
treating the time offsets as independent and ignoring their weak interaction through the set of $T_{0j}$.
The fact that the $T_{0j}$ vary only slightly from event to event further justifies this.
This yields a set of $N_{\mbox{\tiny PMT}}$ independent equations and $N_{\mbox{\tiny PMT}}$ variables, thus,
for each PMT $i$,
\begin{equation}
\nonumber
\frac{d \chi^2}{dt_{0i}} =
\sum_{j\in\{S_i\}} 2 \left( t_{0i} - U_{ij} \right) \, .
\end{equation}
Setting that to zero,
\begin{align}
\nonumber
0 &= \sum_{j\in\{S_i\}} \left( t_{0i} - U_{ij} \right)
% \\
%  &= \sum_{j\in\{S_i\}} t_{0i}
%  - \sum_{j\in\{S_i\}} U_{ij} \, .
\\
\nonumber
 &= t_{0i} N_i
 - \sum_{j\in\{S_i\}} U_{ij} \, .
\end{align}
where $N_i$ is the number of events for which PMT $i$ is in the Select Set.
Then
\begin{equation}
\label{5.3}
t_{0i} = \frac{1}{N_i} \sum_{j\in\{S_i\}} U_{ij} \, .
\end{equation}
As with $T_0$, the $t_{0i}$ can be obtained through an iterative approach, thus:
\begin{enumerate}
\item Process a set of events, finding tracks as described in \S~\ref{sec4} using the current set of offsets,
and accumulating the $N_{\mbox{\tiny PMT}}$ sums in Eq.~\eqref{5.3} as each track is found.
\item Obtain new offsets $t_{0i}$ for each PMT through Eq.~\eqref{5.3}.  Only hits in the Select Set are used for this.
The new offsets become the current set of offsets.
\item If the average change in the offsets is above some pre-set limit, then go back to Step~1.
\end{enumerate}

A typical data set, large enough for use in accurately obtaining a set of offsets, consists of $10^7$ events or more,
from which about $10^6$ tracks are reconstructed in the case where the RPC is required.
Track-finding (Step~1) for such a data set takes several days of processing time on a single processor.
Because the convergence here is quite slow, taking about 100 iterations before Step~3 falls through,
reprocessing the same data as in Step~1 above for each iteration is impractical.

Instead, a minimal set of information describing each track is recorded in an event list.
Whether the tracks are reconstructed with IWS and OWS separately or combined, this minimal set
of information is split by detector and stored separately for the IWS and OWS,
with only IWS hits and $T_0$ being stored in the IWS event list, etc.
With the requirement that both IWS and OWS hits are on a track, each track will contribute one entry to each of the two lists.
No further connection is needed between these two event lists.
% This minimal set includes the tracks's $T_0$ (IWS or OWS), and,
% for each of the track's hits in the Use Set, the hit's PMT index, weight,
% a flag indicating whether it is in the Select Set or not, and the quantity $\tau_{ij} - T_{0pij}$.
% It does not include the trajectory or any other information.
% Given the set of hits from a track, their times, the PMT index (from which the PMT position is available),
% and the track's $T_0$, the full trajectory could be rapidly recovered, but it isn't needed.
At the end of data processing, the event list is used in the iterations over events, replacing Step~1 above with this:
\begin{enumerate}
\item Process the minimal event lists separately for each detector, using the current set of offsets.
Obtain new $T_{0j}$ for each event $j$, iterating through Eqs.~\eqref{3.7}, \eqref{3.8}, \eqref{3.9}.
In obtaining $T_{0j}$ for event $j$, all hits in the Use Set are used, and the hits' square charges used as weights,
as during track-finding, since this step deals with single events.
Accumulate the $N_{\mbox{\tiny PMT}}$ sums in Eq.~\eqref{5.3}.
\end{enumerate}
Steps~2 and 3 remain the same.
This does not permit some hits to be discarded and new hits added as the offsets are improved,
or even for some tracks to be rejected and new ones found, as would happen under full track reconstruction.
For small changes in the offsets these are small effects.
The gain in processing speed with the new Step~1 is essential.
What is given up by not performing the full reconstruction in each iteration is regained by simply repeating the whole
process, including the initial full track-reconstruction before obtaining new offsets with the minimal event lists.
It is generally adequate to repeat this two or three times, though in EH3 the initial offset determination required about 20
passes because of the lower statistics there.
% This need for extra passes in EH3 was driven entirely by the four IWS floor PMTs in the center of the pool,
% which have the lowest statistics of any PMTs, presumably because they are the most poorly illuminated by direct light from muons
% that are also seen by a sufficient number of other PMTs.
Subsequent offset-determinations, using this first set as a starting point, requires no more than two or three passes,
even for EH3, which obtains a precision of better than 0.10~ns in the PMTs with the poorest statistics.\footnote{
  The results presented here obtained a precision of 0.01~ns, primarily to demonstrate that the
  method is stable, and not given to the offsets' gradually ``walking away''.
  This required about 10 passes for EH1 and EH2 (20 for EH3),
  as opposed to the two or three needed to obtain a precision of 0.10~ns.
}

An additional gain in speed is accomplished by breaking the data into small sets and processing them in parallel on,
say, ten processors, reducing the time required for track-finding and generating
partial event lists from $10^7$ events to a few hours.
While more processors are available, using more produces diminishing returns since they share the
same disk system, which ultimately limits the gain in speed.
Offset finding is then done with the combined set of these partial event lists,
which takes only a few minutes on a single processor.

Because this method obtains the offsets for each PMT independently of the others,
it is important to place a common constraint on the offsets.
To accomplish this, after each iteration the weighted-average change in the new offsets (for a given detector)
is subtracted from each of the new offsets, with the result that the average of the offsets is maintained at zero.
The weights used here are simply the numbers of entries for each PMT,
i.e., the number of times a PMT was in the Select Set.
These weights reflect the size of the effect each PMT has on the new offsets.
It is observed that the sequence of new offsets from each iteration diverges if this average is not so-weighted.

Without this constraint of holding the weighted average change to zero,
each iteration is observed to introduce a small, net drift in each of the offsets, uncorrelated between PMTs.
As a plausible explanation for this drift, recall that the stored $T_0$ values pertain to the initial set of offsets.
Even though the $T_0$ are recalculated for each new set of offsets,
the observed net drifts in the offsets produce corresponding drifts in the $T_0$
which propagate into the $t_{0p}(t)$ from Eq.~\eqref{3.7}, yielding an increase in the residuals $\delta t$ from Eq.~\eqref{3.8}.
As this drift accumulates from iteration to iteration,
the $T_0$ values are increasingly shifted, with the result that the iterations produce a sequence of offsets which diverge.
Although this divergence does not necessarily render the offsets invalid,
it certainly makes it impossible to define a convincing termination criterion.
The constraint of holding the weighted average to zero corrects for this.

When the offsets are obtained and Step~3 falls through, the final offsets are corrected so that their {\em unweighted} average
is zero before being stored in the database.
This procedure of correcting the offsets so that their unweighted average is zero is essential
in comparing offsets from different sources.
For example, before the offsets were determined, the 28~ns transit time difference between Hamamatsu and EMI PMTs
was used to define an initial set of default offsets for each detector (IWS and OWS) in each hall, with
-10.182~ns for all of the Hamamatsu PMTs and 17.818~ns for all the EMI PMTs in the IWS in EH1.
These values vary slightly among detectors and halls because it depends on the number of each kind of PMT,
and even varies with time because failed EMI PMTs have been replaced with Hamamatsu PMTs
during the upgrade from six to eight ADs.
Likewise, when comparing offsets obtained from LED calibrations in the early runs, when most of the LEDs were still functioning,
this shift to zero-average is also necessary.
As the LEDs have progressively failed over time, obtaining offsets from muon data has become the only available option.

Some considerable effort was expended in trying to use this method to recover the PMT offsets from the MC data.
Starting with all-zero offsets (i.e., deliberately incorrect offsets), the method produced
satisfactory results for the first four passes, with the number of reconstructed tracks increasing with each pass,
and with each successive set of offsets approaching the ones used by the MC simulation itself.
After the fifth pass, however, the number of reconstructed tracks began to slowly decrease with each successive pass,
indicating that the new offsets were not quite as good as the previous set.
This is never observed with real data.
By the sixteenth pass, the offsets of two poorly illuminated OWS PMTs began diverging without limit.
Again, this is never observed with real data.
There are two likely contributors to these problems, both unique to the MC data.
First, and most important, the simulations produce only about half the PMT charge as is seen in real data.
This is a result of some overly conservative assumptions in the simulations, such as PMT quantum efficiency.
As this had no effect on the primary physics goals of the Daya Bay experiment, no attempt was made to correct this.
Second, the MC data set is only about half the size of a typical set of real data used to determine offsets.
The criteria used to terminate a sequence of passes on a given set of real data are therefore too strict for the MC data,
given these limitations in the MC data set, but creating special termination criteria for MC data
would have defeated the purpose of this effort.
It should also be mentioned that the simulations were made using the original design of PMT placement,
which is slightly different than the as-built placement.
When handling real data, the as-built placements were used, while when handling MC data, the
design placements were used.
This of itself should not cause any difficulties, but it does deserve mentioning.

% A new set of offsets will cause some hits that were used by a track to no longer be used,
% and some hits that were not used by a track to now be used,
% so it is necessary to repeat the whole process of track-finding and
% offset-determination if the offsets were changed by more than about 0.1~ns, if optimal performance is desired.
% In the worse case of starting with all-zero offsets, this generally requires two or three additional passes,
% after the first charge-only pass.
% It takes about 12 passes to get the offset changes all down below 0.001~ns, but this is not really necessary.
% Generally, offsets are determined once per week.
% The offsets from a previous run are essentially unchanged, and a single pass is sufficient to verify this.
%
%

\section{Timing Distributions}
\label{sec7}
 \begin{figure}
 \includegraphics[width=8.5cm]{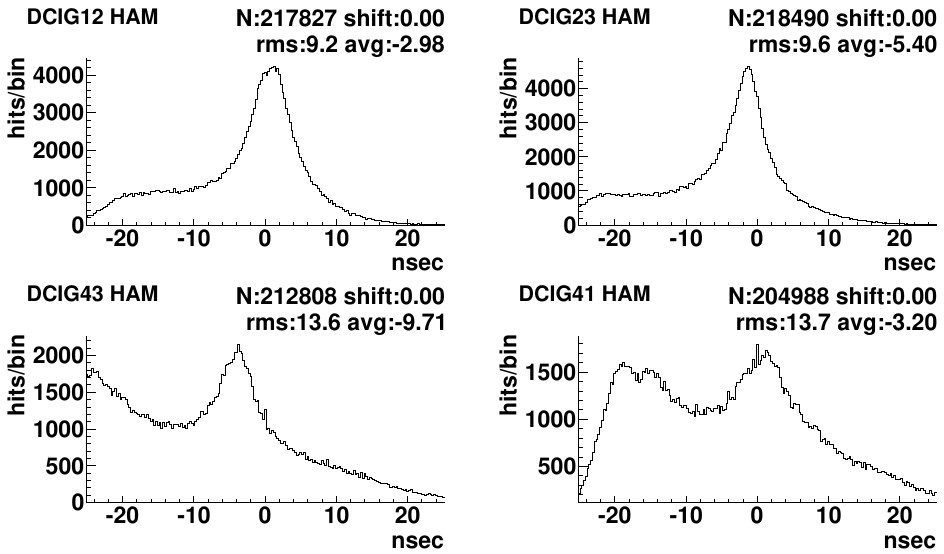}
 \caption{\label{Fig9}
  Distributions of $\delta t$ from Eq.~\eqref{3.8}
  for four typical EH1 IWS PMTs, using all-zero offsets,
  with $Q_{\rm sumMin}=0$ and $Q_{\rm hitMin}=0$.
  The text in the upper right of each plot displays ``N'', the total number of histogram entries
  including underflows and overflows;
  ``shift'', the amount by which a PMT's $t_0$ will shift in ns
  (zero at this point, since the offset-finding algorithm has not yet run);
  and the RMS and average of the distribution in ns,
  including underflows and overflows.
  Because of memory limitations, only half of the 4.5~million tracks reconstructed from IWS, OWS, and RPC data
  in a data set consisting of 38~million events assembled from 99~million detector triggers are included here.
}
 \end{figure}
 \begin{figure}
 \includegraphics[width=8.5cm]{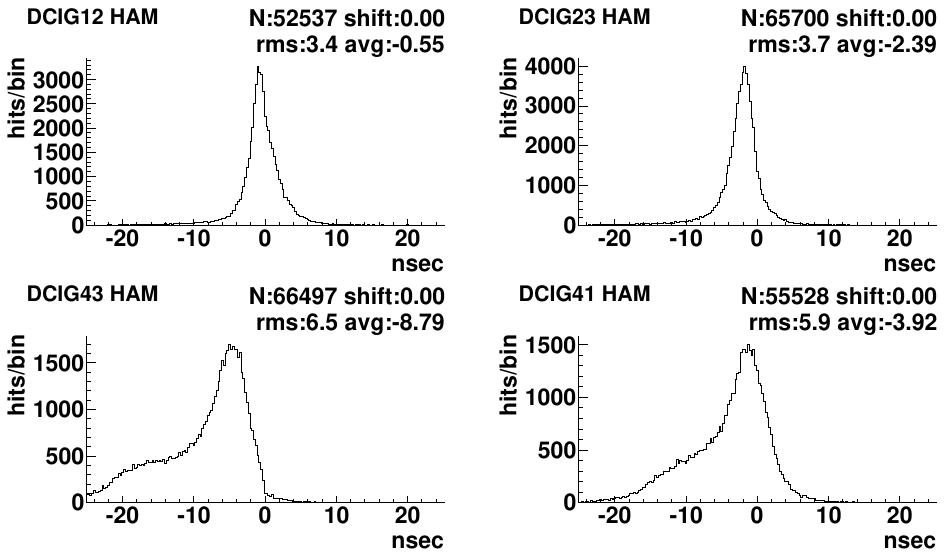}
 \caption{\label{Fig10}
  Same as Fig.~\ref{Fig9}, but with $Q_{\rm sumMin}=300$~pe, $Q_{\rm hitMin}=50$~pe,
  and using the full data set, in which 3~million tracks were reconstructed.
}
 \end{figure}
 \begin{figure}
 \includegraphics[width=8.5cm]{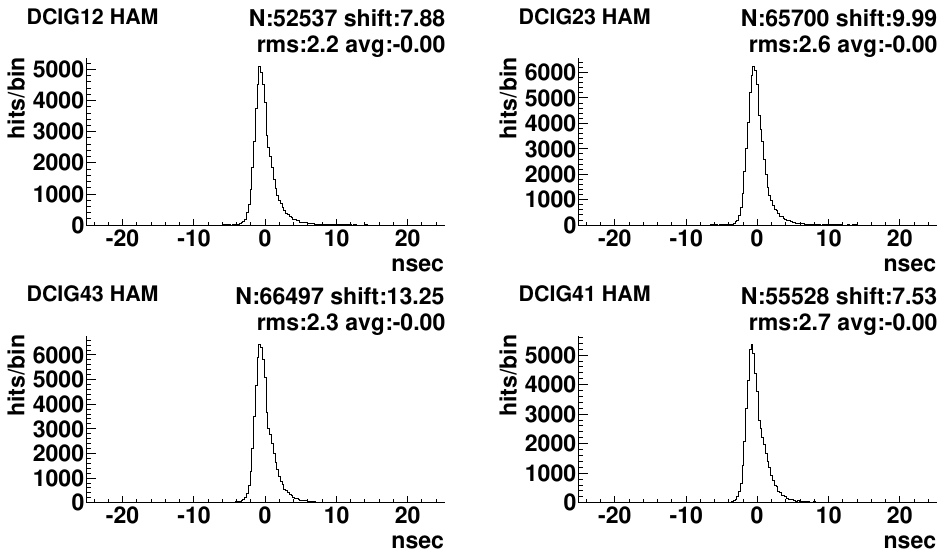}
 \caption{\label{Fig11}
  Same as Fig.~\ref{Fig10}, but using offsets as determined using the charge-only method.
  These plots were produced at the end of the same analysis that produced Fig.~\ref{Fig10} at the beginning,
  before the offsets were determined,
  and show the effect of the iterations determining the $t_{0i}$ in Eq.~\eqref{5.3}.
  The shift in each offset is indicated by ``shift'' in the text at the upper-right of each plot.
}
 \end{figure}
 \begin{figure}
 \includegraphics[width=8.5cm]{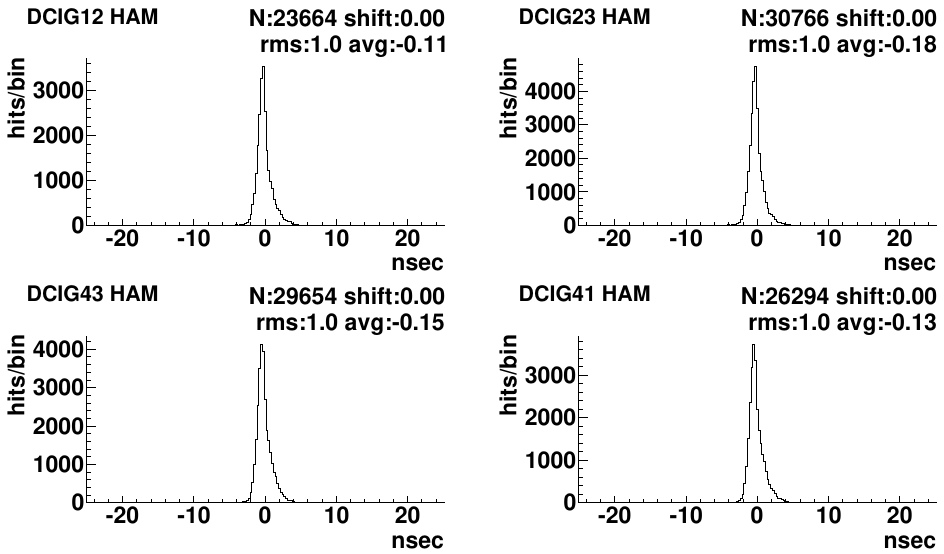}
 \caption{\label{Fig12}
  Same as Fig.~\ref{Fig11}, but now using PMT times to find tracks.
  The offsets used were the same as those used in Fig.~\ref{Fig11}.
  The more than 50\% loss in statistics is due to the more stringent
  requirements of a hit being used by a track when the times are used.
}
 \end{figure}
 \begin{figure}
 \includegraphics[width=8.5cm]{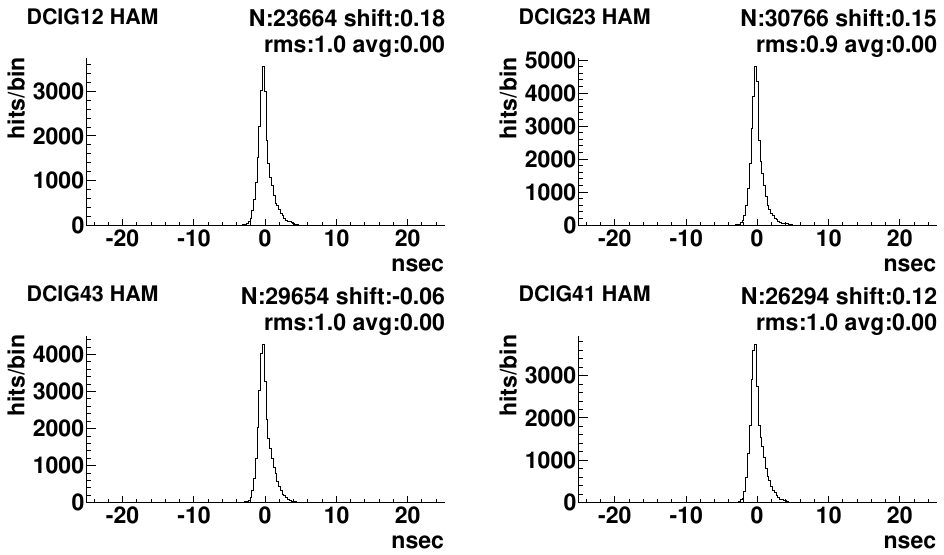}
 \caption{\label{Fig13}
  Same as Fig.~\ref{Fig12}, but after the offsets have been redetermined using the PMT times.
  The ``shift'' values show the change in each PMT's offset.
}
 \end{figure}
 \begin{figure}[!tbp]
   \centering
   \includegraphics[width=0.5\textwidth]{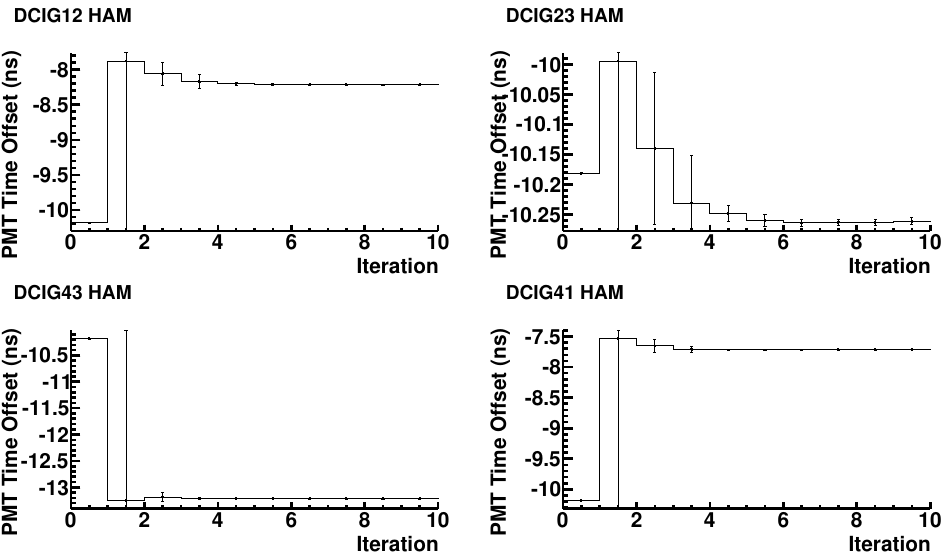}
 % \hfill
 % \begin{minipage}[b]{0.2\textwidth}
 %   \includegraphics[width=0.9\textwidth]{MuonTrackingFig14a}
 % \end{minipage}
 % \hfill
 % \begin{minipage}[b]{0.2\textwidth}
 %   \includegraphics[width=0.9\textwidth]{MuonTrackingFig14b}
 % \end{minipage}
 % \hfill
   \caption{\label{Fig14}
    Time offset $t_{0}$ iterations for four typical EH1 IWS PMTs.
    The level at iteration 0 is that of the default offset.
    The first iteration is that of the charge-only method.
    The next seven are from the ``refined method'', which uses PMT times.
    This demonstrates the convergence of the method.
   }
 \end{figure}
Figures \ref{Fig9} and \ref{Fig10} show the residuals $\delta t$ from Eq.~\eqref{3.8}
for hits in the Select Set using the charge-only method for four typical PMTs,
for two values of $Q_{\rm sumMin}$ and $Q_{\rm hitMin}$.
These plots demonstrate the need for non-zero values for $Q_{\rm sumMin}$ and $Q_{\rm hitMin}$.
Figure \ref{Fig11} shows these residuals after the offsets were determined with the charge-only method.
Figure \ref{Fig12} shows these residuals with the full method using PMT times, with the same offsets used in Fig. \ref{Fig11}.
Figure \ref{Fig13} shows these residuals after the offsets were determined with the full method using PMT times.
While Fig. \ref{Fig13} does not look much different than Fig. \ref{Fig12}, the small shifts of about 0.15~ns have
a significant effect in that the next pass finds about 10\% more tracks.
Figure \ref{Fig14} shows the results of successive ``full'' iterations for four typical PMTs,
where each ``full'' iteration involves one full
track-finding pass followed by a set of iterations over Steps 1-3 as described above.

\section{Tracking with the ADs}
\label{sec8}
It is simple to demonstrate that the earliest light in a scintillator follows the same rule as does Cherenkov light,
even though scintillation light is essentially isotropic.
Using that principle, together with knowledge of the index of refraction of the scintillator in the Daya Bay ADs,
it is straightforward to extend the water pool tracking algorithm to treat the ADs.
Indeed, the problem is much simpler in the ADs, where there are no optical obstacles of any consideration.

\section{Conclusion}
\label{sec9}
A general method of tracking with light detected by PMTs in an optically cluttered environment is described.
Simulated data have been used to determine the method's degree of veracity.
In the analysis of simulated data, the reconstructions were compared only with the simulated primary particles,
which were exclusively high energy muons.
The good agreement there indicates that the tracks reconstructed from real data are almost exclusively high energy muons.
The analysis of real data with this tracker verifies that the simulated muon fluxes used by the Daya Bay experiment are
good approximations to the actual muon fluxes.
While this was demonstrated in Ref.~\cite{MuonPaper}, the comparison there was made with the RPC and RPC Telescopes,
which required some corrections because of the limited acceptance of the RPC Telescopes.
The comparison described here involves no corrections,
and is in good agreement with the muon flux comparison in Ref.~\cite{MuonPaper}.

The ratio of tracks reconstructed from real data to the total number of IWS triggers (Fig.~\ref{Fig7}) is
0.56 for EH1 and EH2, but only 0.25 for EH3.
It is not clear how much of this difference is due to overburden and how much is due
to the different sizes and geometries of the three halls, since EH1 and EH2 are essentially identical,
and with very nearly the same overburden, while EH3 is about twice the size of the others, and with more
than three times the overburden.
There are indications of showers from multiple-track reconstructions and from events where light is seen by many PMTs
but where no tracks could be reconstructed.
Because it is not obvious how much of this is due to showers,
and how much is due to too much reflected light and not enough direct light,
it is only possible to give limits regarding showers.
From the ratios given above, fewer that half of all IWS triggers in EH1 and EH2, and three quarters in EH3,
probably correspond to showers.

It is important to emphasize that this tracker is not used in any of Daya Bay's published physics results
at the time of this writing.
The tracker's role so far has been only to independently verify the simulations of cosmic muon fluxes.
It in no way contributes to the vetoing function of the water pools,
which is indistinguishable from being 100\% efficient~\cite{MuonPaper}.
Furthermore, none of Daya Bay's physics results depend on the determination of the water pool PMT timing offsets.
Only the tracker described here presently makes use of these offsets.
However, the technique described here for determining these offsets may have some general application in other experiments.

An ongoing study of cosmic muons with this tracker will attempt to verify the realism of simulated data being used
in an independent effort to determine neutron production at Daya Bay.
There is also an ongoing effort to reconstruct atmospheric showers producing coincident muons in two or three of Daya Bay's
experimental halls.
\appendix
\section{The Event Display}
\label{AppA}
 \begin{figure*}
 \centering
 \includegraphics[width=\textwidth]{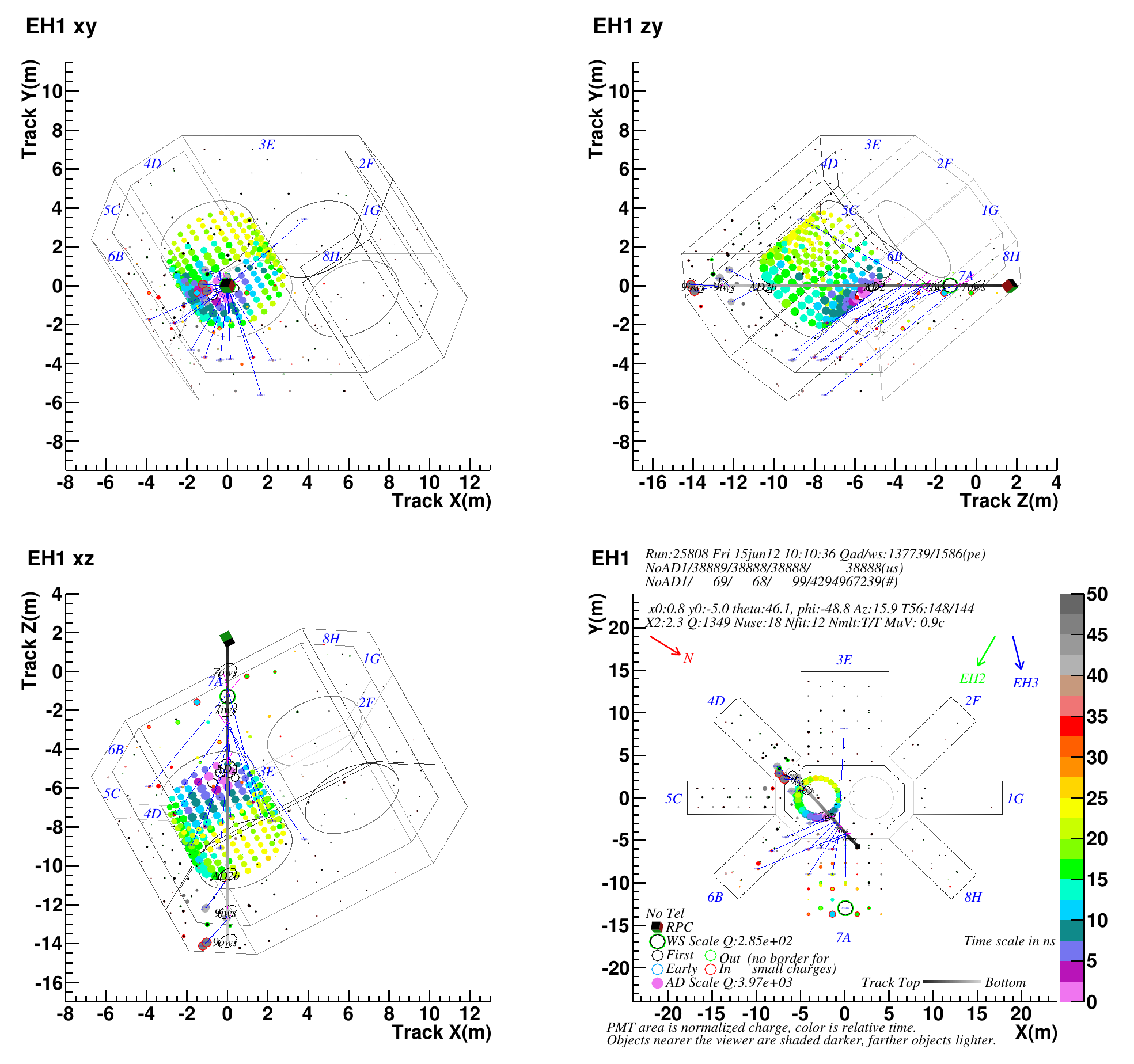}
 \caption{\label{Fig15}
    The event display (see text), showing a track reconstructed with PMT data from the IWS and OWS.
 }
 \end{figure*}
An event display was developed for viewing a sample of the reconstructed tracks, and this was of invaluable assistance.
A typical event in EH1 with a reconstructed track is shown in Fig.~\ref{Fig15}.
This display contains a wealth of information, both graphical and textual.
Some of the information, such as PMT labels, can only be seen by zooming in on the image.
Such small text keeps the displays relatively uncluttered and makes it possible to identify which label belongs to which PMT.
The main title, displayed in each of the four views, indicates the hall, in this case EH1.
The graphical information includes:
\begin{enumerate}
\item A projection looking straight down the reconstructed track (track $z$), onto the track's $xy$ plane.
\item A projection onto the track's $zy$ plane.
The $y$ axis aligns with that of the track $z$ view.
\item A projection onto the track's $xz$ plane.
The $x$ axis aligns with that of the track $z$ view.
\item A hall $xy$ view, where the walls are splayed out for a complete view of the pool's walls.
\item When there are no tracks, the first three views are projections onto the hall's $xy$, $zy$, and $xz$ planes.
\item The ADs are visible in each view.
In the early runs with only six ADs, the missing ADs in EH2 and EH3 are not displayed.
\item The walls are all labeled with their two designations (i.e., engineering and software, see Fig.~\ref{Fig1}), as ``1G'' etc.
In the three non-splayed views, these labels appear at the top of the pool, where $z=0$.
This is important to keep in mind since the top of the pool is not always closest to the viewer.
\item PMT data are represented by circles, where the charge is represented by the area.
The area of the largest circle is fixed, and corresponds to the PMT with the largest charge.
The other PMTs' charges are normalized by the largest charge, done separately for
the pool and the ADs.
\item PMT times are represented by colors, according to the color scale on the right of the splayed view.
When a detector has hits on a track, the earliest hit on the track sets the zero-time on the color scale.
When a detector does not have any hits on a track, this zero-time is set to the earliest of the detector's hits
with a charge of at least 1~pe.
The latest times are off-scale, and are depicted in black.
The color scale has been chosen so as to make identification of PMT times easier
than would be possible with a more gradual color scale.
\item Only PMTs that are identified as belonging to a track are displayed, whether or not they are in the Use Set.
PMT labels are displayed only for those on a track and in the Use Set.
\item If a detector is not used in tracking, all of its PMTs (in the readout) are displayed,
but these are not labeled in order to avoid excessive clutter.
\item Because of the order in which the various objects are drawn, some items, such as PMT labels, may be obscured in some views.
The display does not attempt to draw objects with regard to distance from the viewer, so an IWS PMT may appear over an AD PMT,
even though the IWS PMT might be behind the AD.
\item An RPC hit is depicted as a small cube, with the hall-$z$ faces colored black, the $x$ brown, and the $y$ green.
\item An RPC Telescope hit is depicted as a small cube, with the hall-$z$ faces colored gray, the $x$ magenta, and the $y$ cyan.
These are relatively rare, because of the small size of the Telescope.
\item North is indicated by a red arrow in the upper left of the splayed view and the letter ``N''.
The directions to the other two halls is indicated with green and blue arrows in the upper right of the splayed view,
labeled with the hall designations.
\item A symbol key on the lower left of the splayed view includes:
\begin{itemize}
\item[{\bf Tel}] If there are RPC Telescope data in the display, the symbol for a Telescope hit is shown.
Otherwise ``No Tel'' appears.
\item[{\bf RPC}] If there are RPC data in the display, the symbol for an RPC hit is shown, otherwise ``No RPC'' appears.
\item[{\bf WS Scale}] The IWS or OWS PMT with the largest charge is displayed, along with the color corresponding to its time.
The charge on that PMT is given in units of pe.
If there is no IWS or OWS in the readout, ``No WS'' appears.
\item[{\bf First}] This shows the earliest PMT on the track, where white corresponds to the zero-time, which necessitates
drawing it as a white-filled black circle.
\item[{\bf Early}] A PMT not on the track may be earlier that the zero-time,
in which case it is represented by a blue-outlined circle.
\item[{\bf AD Scale}] The AD PMT with the largest charge is displayed, along with the color corresponding to its time.
The charge on that PMT is given in units of pe.
If there is no AD in the readout, ``No AD'' appears.
\item[{\bf Out/In}] This is peculiar to the OWS, with outward-facing PMTs indicated with a thin green
border, and inward facing PMTs with a thin red border.
This border does not appear on IWS or AD PMTs, where they all face inward,
nor for OWS PMTs with a small charge.
\end{itemize}
\item The symbol labeled ``Track Top'' and ``Bottom'' at the lower right shows that a track is shaded differently
along its length to make it easier to visually distinguish the track's direction.
This concept is followed by all the fixed lines in the display, with darker tones appearing closer to the viewer, and
lighter tones farther, as briefly mentioned in the text just below the splayed view.
\item Small ellipses indicate intersections of the track with various surfaces.
These are circles projected onto a surface,
and labeled to indicate the surface, e.g., ``9iws'' for the floor of the IWS, ``AD2'' for the side of AD2
(there are two of these if the track both enters and leaves through the sides),
``AD2b'' for the bottom of AD2, and so on.
\item The blue lines are the light paths connecting points $p$ and $q$ in Fig.~\ref{Fig2}, for hits in the Use Set.
None of these ever intersect any surfaces, else they would simply be absent.
\item Some of the blue lines are extended with a short magenta line, representing a positive time residual (i.e., it
would push the track away from the PMT), or overlaid with a short red line for a negative residual.
These red and magenta lines are ``bent'' in the splayed view, because of the nature of the splaying transformation.
\end{enumerate}
 \begin{figure*}
 \centering
 \includegraphics[width=\textwidth]{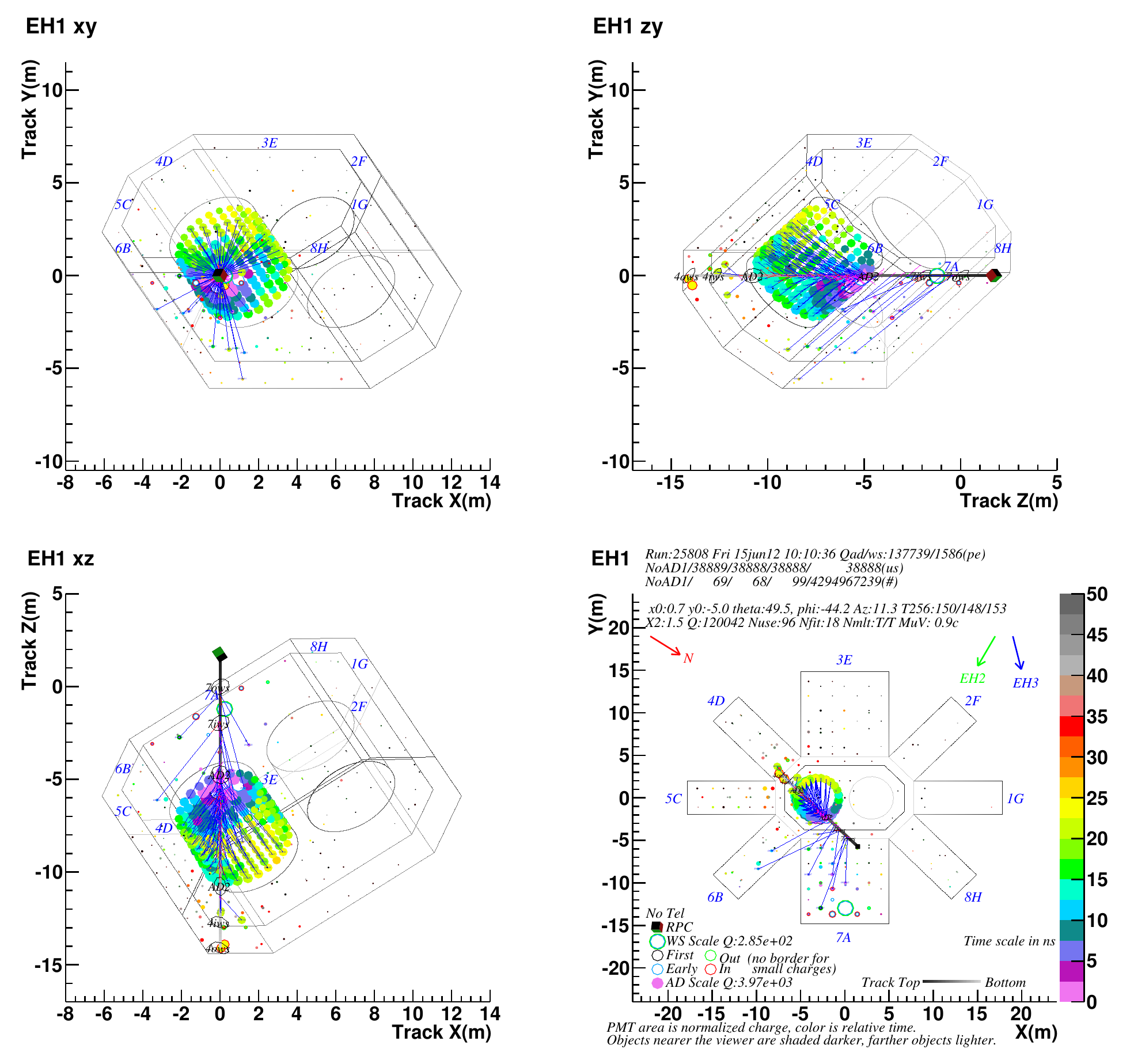}
 \caption{\label{Fig16}
    This shows the same event as in Fig.~\ref{Fig15}, but now with the ADs used for tracking.
 }
 \end{figure*}
The textual information appears mostly at the top of the splayed view, and includes:
\begin{enumerate}
\item Run number, time of the event (China Standard Time),
and the total charge in the ADs (if present in the readout)
and the pool (IWS+OWS), with these two total charges separated by a forward slash and in units of pe,
and labeled ``Qad/ws''.
\item The trigger times for each of the detectors' readouts, separated by forward slashes.
In EH1 (and EH2) these are AD1/AD2/IWS/OWS/RPC, in that order.
If a detector is missing in the readout, it is indicated with, for example, ``NoAD1'', or,
if both ADs were missing, ``NoAD12'', in the interest of brevity.
In EH3, these are AD1/AD2/AD3/AD4/IWS/OWS/RPC.
For the period of running with only six ADs, EH2 omits AD2 and EH3 omits AD4.
These are in units of $\mu$s, as indicated by the ``(us)'' at the end of this line.
The values for the different detectors should be nearly the same, since the reconstruction program is requiring the readouts
to correspond to the same event, as recorded separately by the various detectors.
\item The trigger numbers for each of the detectors, displayed in the same fashion as the times, and
indicated by ``(\#)'' at the end of the line.
The RPC trigger numbers are typically quite large, a testament to their high triggering rate.
In this case it has overflowed and is in the process of appearing to count backward, as seen from the next event
(see the supplemental material).
It is likely that the RPC trigger counter simply failed to reset at the beginning of this run,
a thing of no serious consequence if true.
\item If a track is reconstructed, its basic parameters $x_0$, $y_0$, $\theta$ and $\phi$ are shown to limited precision
and labeled with ``x0'', ``y0'', ``theta'', and ``phi'',
along with the azimuth (compass heading) of the track, labeled ``Az''.
The angles are expressed in degrees.
At the end of that line, $T_0$ is displayed for each of the detectors used by the track, labeled in abbreviated fashion with,
in this case,
``T56:148/144'', which means these are the $T_0$s for each of the detectors with ID 5 and 6,
corresponding to the IWS and OWS.
\item On the last of these lines is given the $\chi_\nu^2$ of the track, labeled ``X2''.
Next is the total charge of hits on the track, labeled ``Q''.
Then comes the number of hits in the Use Set, labeled ``Nuse''.
That is followed by the number of iterations in the reconstruction process, labeled ``Nfit''.
Next appears the IWS and OWS multiplicity triggers, separated by a forward slash and labeled ``Nmlt''.
This shows ``T'' for true or ``F'' for false.
\item Finally is the muon velocity as described earlier, labeled ``MuV'' and given in units of $c$.
\end{enumerate}
Tracking in the AD is switched off in Fig.~\ref{Fig15}, but it is quite apparent that the IWS- and OWS-determined track matches
well with the charge and timing distributions in AD2.
Figure~\ref{Fig16} shows the same event, but with the ADs used for tracking.
The reconstruction is improved somewhat, with $\chi_\nu^2$ reduced to 1.5 from 2.3, but the trajectory is not much different,
The number of iterations required increased from 12 to 18,
and the selection of IWS/OWS hits employed by the track changed somewhat.
The supplemental materials include a file showing 40 events with the ADs not used in tracking,
and another showing 10 events with the ADs used in tracking.
Both of these files were generated from the same group of raw events.

$$
$$
\section{Acknowledgments}
I would like to thank
B. Viren and D. Jaffe
for informative discussions,
and L. Whitehead for her careful reading and long list of most helpful comments.
I would also like to thank the Daya Bay reviewers L. Zhan and M. Wang for their comments,
and J.P. Ochoa-Ricoux for his penetrating questions which caught a few errors.
This work was supported by the United States Department of Energy, under Contract No. DE-AC02-98CH10886.
%\newpage

%els \bibliographystyle{elsarticle-num} %els
\bibliography{MuonTracking}

\end{document}